\documentclass[a4paper,11pt]{article}
\usepackage{jinstpub} 
\usepackage{lineno}

\title{\boldmath A two-stage time-stretching TDC with discrete components}

\author{Yanbo Chu}
\author[1]{and Zhicai Zhang \note{Corresponding author.}}
\affiliation{Department of Engineering Physics, Tsinghua University, \\
Beijing 100084, China}

\emailAdd{zhicaizhang@tsinghua.edu.cn}

\abstract{This paper presents the design and testing of a time-stretching-based time-to-digital converter (TDC) implemented with discrete components. The TDC utilizes capacitor charging and discharging to achieve a time resolution of under 100 ps using a 100 MHz clock counter on a low-power, low-cost FPGA, achieving a time amplification factor of over 100. A two-stage time-stretching architecture is employed to reduce the conversion time to below 300 ns for a 10 ns input range. An onboard calibration system, including a pulse generation circuit, is implemented, and calibration results are presented. This system serves as a proof-of-concept platform for circuit optimization toward an ASIC implementation of a front-end TDC targeting future 4D pixel detectors at hadron colliders, with goals of sub-50 ps resolution and power consumption at the $\mu$W/channel level. Additionally, the design offers a modular, low-cost solution for extracting signal arrival times with 100 ps precision in particle physics experiments, such as photoelectron timing extraction for photodetector readout in neutrino experiments.}


\begin{document}
\maketitle
\flushbottom

\section{Introduction}
\label{sec:intro}

Precision timing detectors capable of measuring particle arrival times with resolutions on the order of tens of picoseconds (ps) are critical for high-energy physics experiments. For example, for hadron colliders, precise timing information for charged particles is essential for efficient vertex reconstruction and pileup mitigation under high instantaneous luminosity conditions~\cite{cmsmtdtdr,atlashgtd}; in neutrino experiments, simultaneous reconstruction of particle energy and direction becomes possible when Cherenkov and scintillation light in slow scintillators can be separated via high-precision photon timing measurements at the photodetector~\cite{Luo_2023}.

These applications demand time-to-digital converters (TDCs) with resolutions of tens of picoseconds. While achieving such resolution is not inherently challenging—modern solutions like precision logic gate delays or high-speed waveform sampling exist—the latter approach, commonly used in neutrino experiments~\cite{JNEReadout}, suffers from high power consumption (exceeding $1~\mathrm{W}$ per channel), large data rates (hundreds of bytes per pulse), and prohibitive costs (up to \$1,000 per channel for GS/s ADCs). Such trade-offs are untenable for collider experiments, where channel counts are orders of magnitude higher, and front-end power budgets are severely constrained by cooling limitations. Instead, delay-line-based TDCs, achieving sub-$20~\mathrm{ps}$ resolution in modern CMOS technologies~\cite{ALTIROC,ETROC,ETROCTDC}, are widely adopted. These typically consume $\sim\mathrm{1~mW}$ per channel at a few percent occupancy, suitable for detectors like CMS MTD and ATLAS HGTD~\cite{cmsmtdtdr,atlashgtd}, where pixel sizes ($\sim\mathrm{1~mm^2}$) permit power densities of $\sim\mathrm{1~W/cm^2}$. However, for finer-pixel detectors (e.g., $\mathrm{50~\mu m \times 50~\mu m}$ innermost pixel layers), power must be reduced to the $\sim\mathrm{1~\mu W}$/channel level—far beyond the reach of current delay-line TDCs. In further upgrade of innermost layers of pixel detectors at HL-LHC (beyond the on-going Upgrade II of ATLAS or CMS pixels), or for pixel detectors at future hadron colliders such as FCC-hh, a 4D pixel detector with the timing and power requirements listed above are certainly on the table, with TDCs meeting such requirements yet to be developed.

This paper presents a time-stretching TDC design that addresses this gap. By amplifying input pulse widths via capacitor-based fast charging and slow discharging, the stretched pulse can be digitized with a slower clock. For instance, a $40~\mathrm{MHz}$ counter achieves an effective least significant bit (LSB) of $25~\mathrm{ps}$ with a time-stretching factor of 1000. With $\mathrm{nA}$-level charging currents and tens of $\mathrm{MHz}$ clocks, such TDCs can potentially reduce power to $\sim\mathrm{1~\mu W}$ per channel, making them ideal candidates for 4D pixel detectors in future colliders.

The primary limitation of time-stretching TDCs is their dead time, which scales with the stretching factor. For high-rate environments like the Large Hadron Collider (LHC), where innermost pixel hit rates reach $\sim100~\mathrm{kHz}$~\cite{CERN-LHCC-2017-021}, dead times must remain below $1~\mathrm{\mu s}$ to maintain high efficiency. A single-stage stretcher with a factor of 1000 would limit the input range to $1~\mathrm{ns}$—impractical for most LHC applications.

To mitigate this, we employ a two-stage time-stretching architecture. The input pulse is first stretched by a factor $S_0$, measured with a low-speed clock, and the residual tail (under one clock cycle) is stretched again by $S_1$ for a second measurement. The total stretching factor becomes $S_0S_1$, while the dead time reduces to $T_0S_0 + T_1S_1$, where $T_0$ is the input range and $T_1$ the clock period. For typical LHC parameters ($T_0 = T_1 = 25~\mathrm{ns}$), dead time drops to $(S_0 + S_1) \times 25~\mathrm{ns}$—far below the $S_0S_1 \times 25~\mathrm{ns}$ of a single-stage design.

Optimizing such multi-stage TDCs (e.g., stretching factors, clock speeds, calibration) requires iterative prototyping, which is cost-prohibitive in ASIC form. As a proof-of-concept, we implement a discrete-component prototype (using SMT capacitors, transistors, and FPGAs) focused on resolution and dead time rather than power efficiency (achieving $\mathrm{\mu W}$ levels would require CMOS integration). The prototype demonstrates sub-$100~\mathrm{ps}$ resolution and sub-$300~\mathrm{ns}$ dead time with a $100~\mathrm{MHz}$ clock, already meeting the needs of applications like neutrino experiment timing readout. The following sections detail the design, testing, and optimization of this system.

\section{TDC design and simulation with discrete components}
\label{sec:design}

\begin{figure}[htbp]
\centering
\includegraphics[width=.45\textwidth]{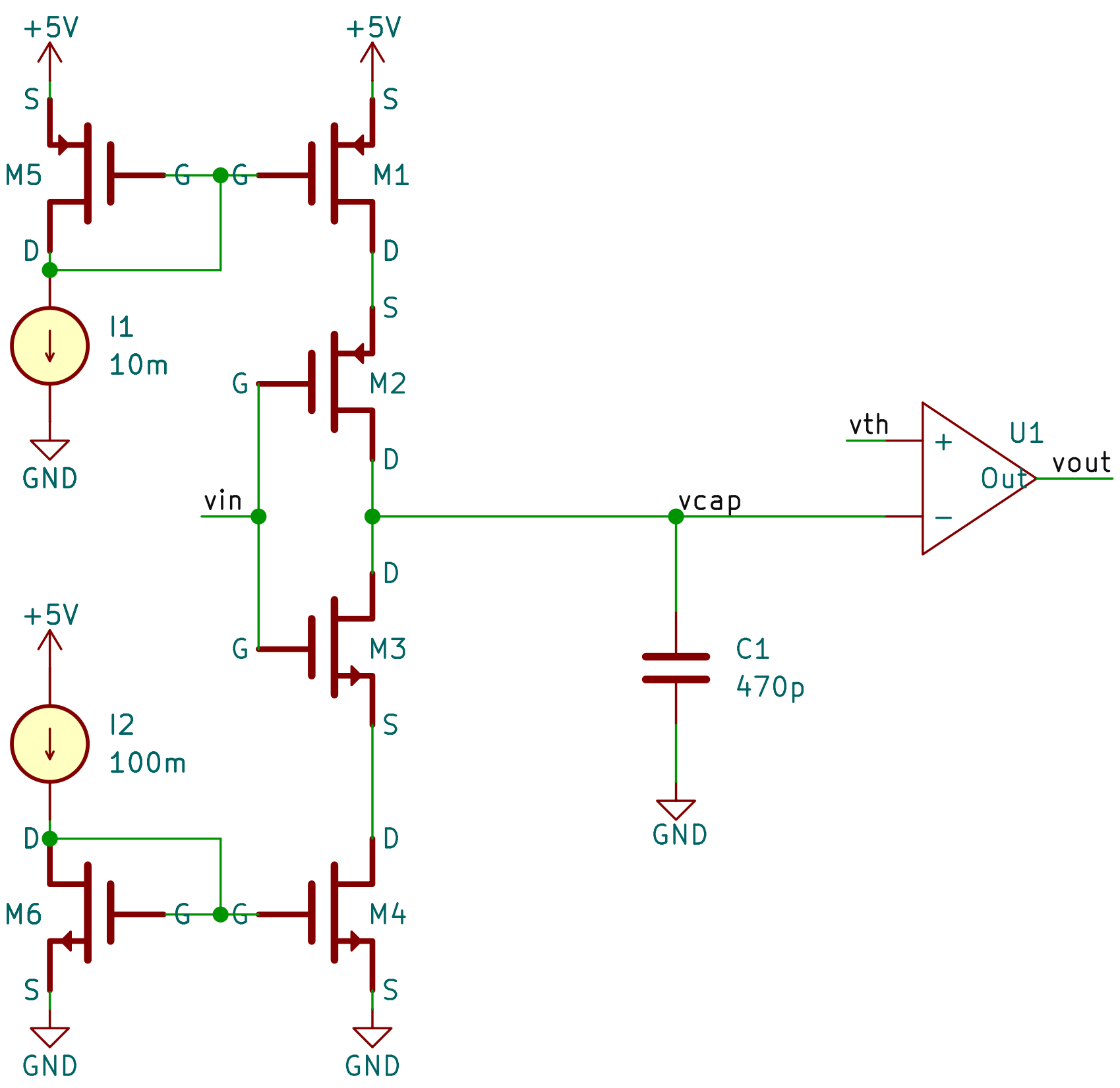}
\qquad
\includegraphics[width=.45\textwidth]{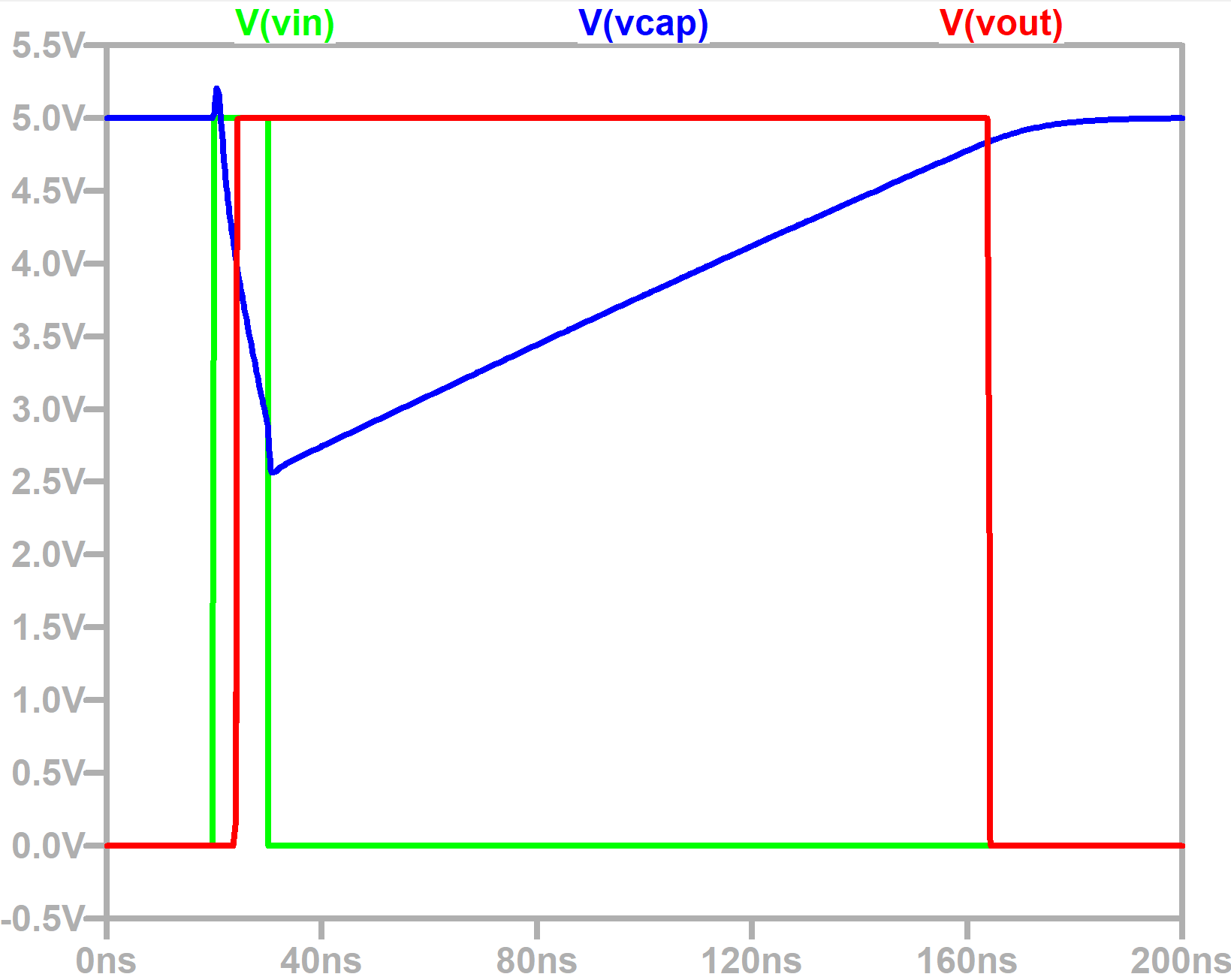}
\caption{Left: schematic of a time-stretching unit with discrete components; right: example pulses of vin, vcap, and vout of the schematic on the left from SPICE simulation (vth is set to $4.8~\mathrm{V}$).\label{fig:circuit_singlestage}}
\end{figure}

The time-stretching unit circuit, shown in figure~\ref{fig:circuit_singlestage}, forms the core of the TDC prototype. When no input signal is present (vin low), PMOS transistor M2 remains ON while NMOS transistor M3 stays OFF, maintaining capacitor C1 (vcap) at its high voltage level ($5~\mathrm{V}$ in this implementation). During an input pulse (vin high), M2 turns OFF and M3 turns ON, discharging C1 through current I2 ($100~\mathrm{mA}$ in this design). This current is mirrored to M3 and M4 via a current mirror formed by M6 and M4. The resulting voltage drop rate across C1 is given by:

\begin{equation}
\label{eq:1}
\bigg|\frac{\mathrm{d}V}{\mathrm{d}t}\bigg|_{\mathrm{on}} = \frac{I2}{C1} = 0.21~\mathrm{V/ns}
\end{equation}

When the input pulse returns to low, M2 switches ON and M3 switches OFF, initiating the charging of C1 through current I1 ($10~\mathrm{mA}$ in this implementation). The resulting voltage rise rate across C1 is given by:

\begin{equation}
\label{eq:2}
\bigg|\frac{\mathrm{d}V}{\mathrm{d}t}\bigg|_{\mathrm{off}} = \frac{I1}{C1} = 0.021~\mathrm{V/ns}
\end{equation}

Equations~\eqref{eq:1} and \eqref{eq:2} demonstrate that the capacitor's charging rate is significantly slower than its discharging rate, with their ratio matching I1/I2. By measuring the time duration during which vcap remains below a defined threshold (the pulse "width"), the output width becomes proportional to both the input width and the current ratio. For an ideal case with the threshold set near $5~\mathrm{V}$, the stretching factor (output-to-input width ratio) approximates:

\begin{equation}
\label{eq:3}
    S = \frac{\mathrm{output~width}}{\mathrm{input~width}} = 1 + \frac{I2}{I1} 
\end{equation}
This yields a stretching factor of 11 for the circuit configuration shown in figure~\ref{fig:circuit_singlestage}, effectively amplifying the input pulse width by a factor of 11. The right plot of figure~\ref{fig:circuit_singlestage} displays representative SPICE simulation results for a $10~\mathrm{ns}$ input pulse, showing both the capacitor voltage waveform and discriminator output. The simulation employs high-speed components including a TLV3601 comparator and NX3008 series transistors in its model.

As noted in section~\ref{sec:intro}, power consumption is not a primary optimization target for this discrete-component prototype. The required currents (I1 and I2) in the time-stretching circuit are determined by the I-V characteristics of the selected transistors. While alternative fast-switching transistors (such as the ALD1106/ALD1107 series) can achieve comparable performance with significantly lower current consumption (typically $< 1~\mathrm{mA}$), the NX3008 series was selected for this prototype based on availability and cost considerations.

\begin{figure}[htbp]
\centering
\includegraphics[width=.9\textwidth]{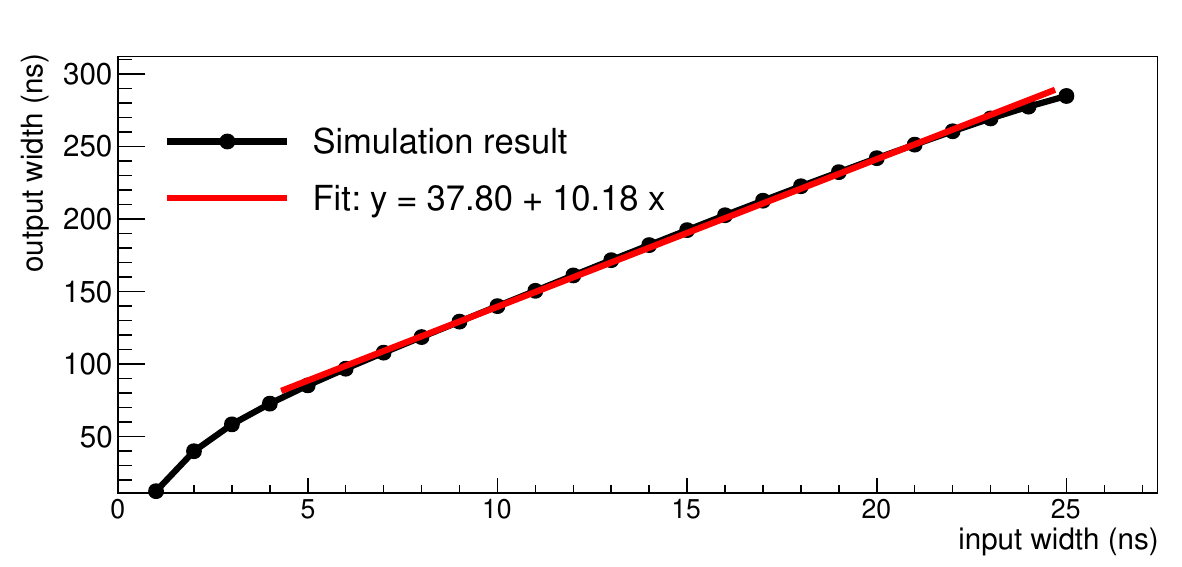}
\caption{ Width of output pulse as a function of the input pulse width of the time-stretching unit from SPICE simulation. The red curve is a fit with a linear function, with the fit result shown in the legend.\label{fig:simulation_singlestage}}
\end{figure}

Figure~\ref{fig:simulation_singlestage} shows the simulated output pulse width versus input pulse width (scanned from $\mathrm{1}$ to $\mathrm{25~ns}$) from SPICE simulations, including a linear fit to the simulation data points. The plot reveals excellent linearity between input and output widths in the $\mathrm{5}$--$\mathrm{25~ns}$ range. For time-of-arrival (TOA) measurements, where the input pulse arrival time is measured relative to a clock rising edge, adding an extra clock cycle (or half-cycle) to the discriminated TOA before width measurement can avoid the non-linear region ($<5~\mathrm{ns}$). For instance, with a $\mathrm{100~MHz}$ clock (10 ns period), the original TOA range of 0--10 ns becomes 10--20 ns after adding one clock cycle, keeping all measurements within the linear region of the time-stretching unit.

\begin{figure}[htbp]
\centering
\includegraphics[width=1\textwidth]{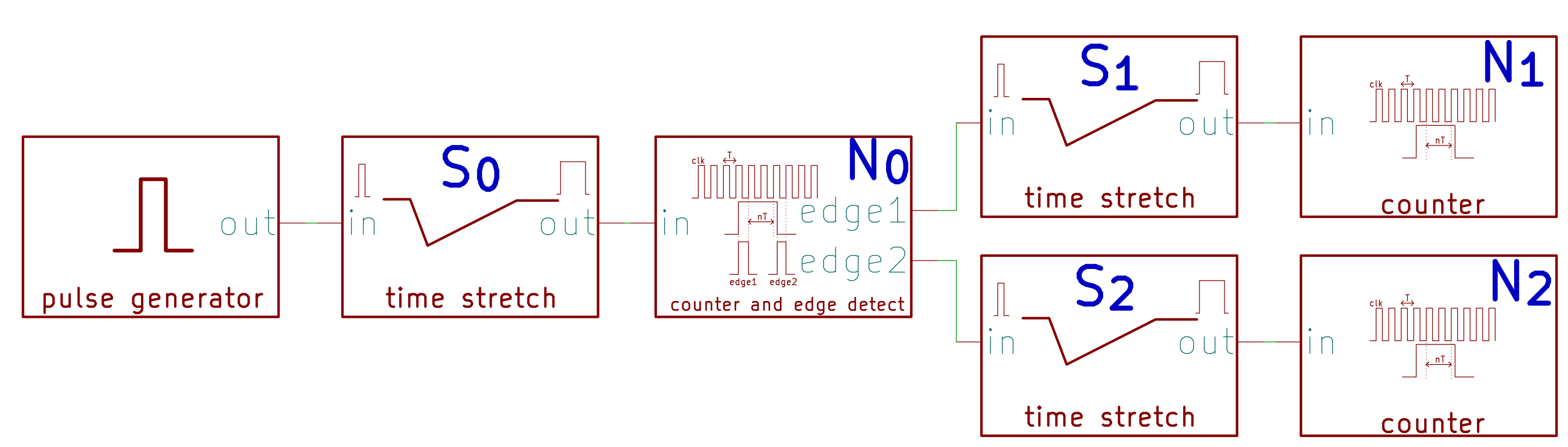}
\caption{Diagram of the entire two-stage TDC based on time-stretching unit, pulse generator, and counters.\label{fig:system_diagram}}
\end{figure}

Once the time-stretching circuit is implemented, constructing a multi-stage TDC becomes relatively simple. Figure~\ref{fig:system_diagram} illustrates the schematic of the implemented prototype. A few-ns-wide input pulse (generated by an onboard pulse generator, described later in this section) undergoes initial stretching by a factor $S_0$, then measured by a $100~\mathrm{MHz}$ counter. This yields $N_0$ – the number of full clock cycles in the stretched pulse, representing the most significant bit (MSB) of the TDC output.

During counting, the rising and falling edges of the stretched pulse are not necessarily aligned with the clock's rising and falling edges. Edge-to-edge detection circuits (made with flip-flops and logic gates) are used to capture the two extra tails of the pulse:

\begin{itemize}
    \item Edge1: pulse rising edge to the next clock rising edge.
    \item Edge2: pulse falling edge to the next clock rising edge.
\end{itemize}

These residual tails (edge1 and edge2) are subsequently processed by two additional time-stretching units with stretching factors $S_1$ and $S_2$, respectively. Following secondary stretching, the pulses are measured using  $100~\mathrm{MHz}$ counters, generating counts $N_1$ and $N_2$ that comprise the least significant bits (LSB) of the TDC.

A low-cost FPGA (Intel Max 10) serves as the counter and edge-detection unit in the prototype shown in figure~\ref{fig:system_diagram}.

The complete time-stretching chain reconstructs the input pulse width as:
\begin{equation}
\label{eq:4}
\mathrm{measured~input~width} = \frac{N_0T + N_1T/S_1 + T - N_2T/S_2}{S_0},
\end{equation}
where $T = 10~\mathrm{ns}$ (clock period). For $S_0 \approx S_1 \approx S_2 = S \approx 10$, the theoretical LSB becomes:
\begin{equation}
\label{eq:5}
\mathrm{LSB} \approx \frac{T}{S^2} \approx 100~\mathrm{ps}.
\end{equation}
This corresponds to a theoretical resolution of $100/\sqrt{12} \approx 29~\mathrm{ps}$, excluding jitter contributions from circuit components (to be quantified experimentally in later sections).

The system dead time can be derived from the stretched pulse duration and subsequent processing. For typical TOA measurements with a $\mathrm{100~MHz}$ clock (10 ns period), the original $\mathrm{10~ns}$ TOA range becomes $\mathrm{15~ns}$ after adding a half-clock cycle offset to avoid non-linearity. This $\mathrm{15~ns}$ pulse stretches to $\mathrm{150~ns}$ when processed by the $S_0=10$ stage. The edge1 and edge2 residuals each require $\leq\mathrm{150~ns}$ processing, resulting in a maximum total dead time of $\sim\mathrm{300~ns}$ – sufficient for both collider and neutrino experiment applications.

\begin{figure}[htbp]
\centering
\includegraphics[width=.50\textwidth]{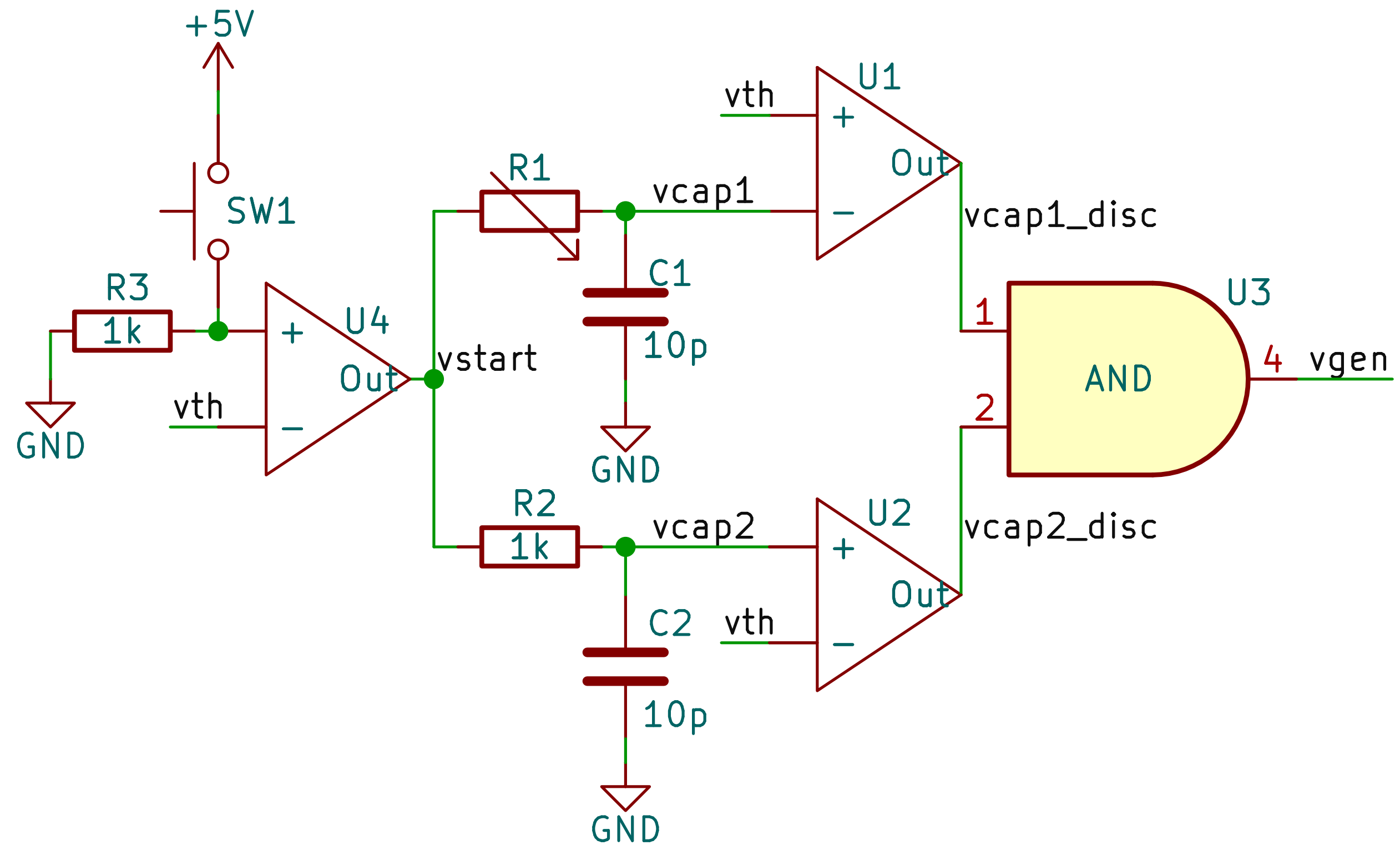}
\qquad
\includegraphics[width=.40\textwidth]{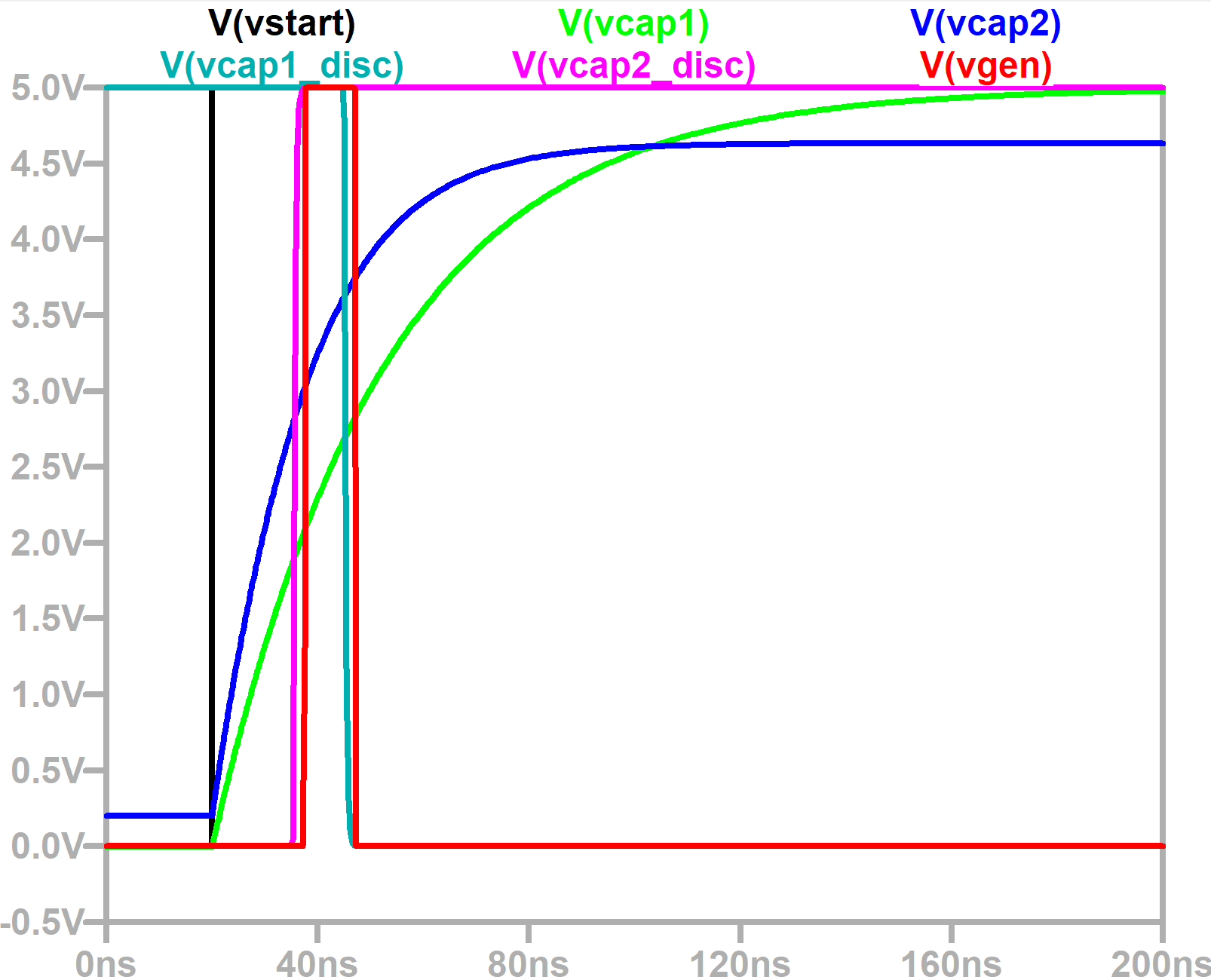}
\caption{Left: schematic of the RC-delay pulse generator used in the prototype; right: example pulses of different nodes of the circuit on the left from SPICE simulation (a $2.5~\mathrm{V}$ threshold voltage vth is used).\label{fig:circuit_pulse_generator}}
\end{figure}

To generate the required $\mathrm{ns}$-level test pulses for the TDC testing, the prototype PCB incorporates an onboard precision pulse generator. This circuit exploits differential RC delays of an input edge (triggered by either a push-button or FPGA-generated signal via a fast comparator), as shown in figure~\ref{fig:circuit_pulse_generator}. Pulse width adjustability is achieved through variable resistor R1. The design employs fast-switching components including LT1711 comparators and an SN74LVC1G08 AND gate. The right panel of figure~\ref{fig:circuit_pulse_generator} displays SPICE simulation results using these component models. An fast driver (LMG1020) is used to deliver pulses to the TDC circuitry.

\section{Performance of single-stage time stretching unit}
\label{sec:perf_single}

\begin{figure}[htbp]
\centering
\includegraphics[width=.90\textwidth]{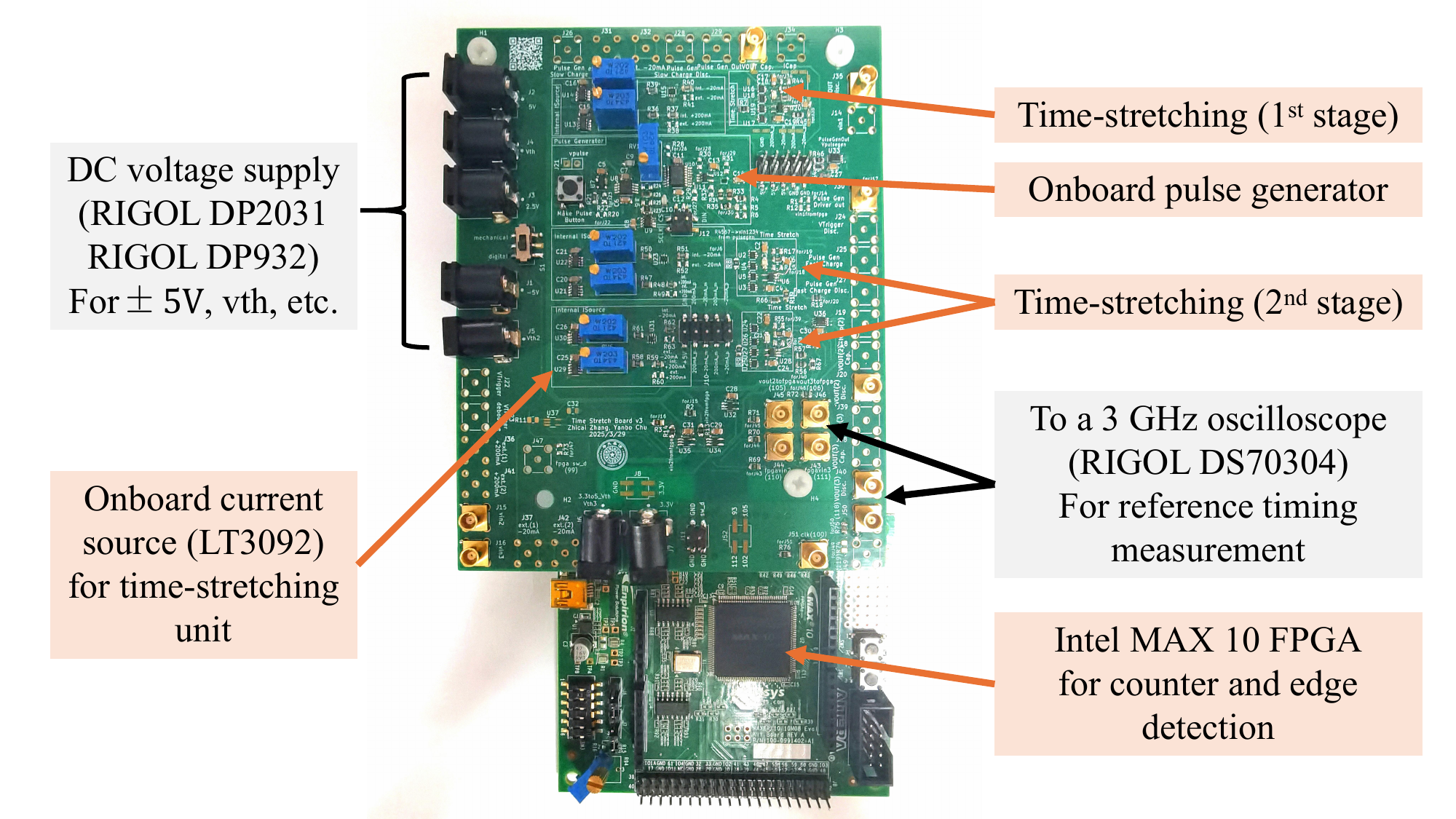}
\caption{Photo of the prototype TDC PCB and the test setup illustration.\label{fig:test_setup}}
\end{figure}

The time-stretching units and pulse generator are integrated on a single PCB. Figure~\ref{fig:test_setup} shows the prototype PCB and test configuration. LT3092 programmable current source regulators, mounted on-board, provide precise currents for the time-stretching circuits. An Intel MAX 10 FPGA evaluation kit interfaces with the PCB via pin headers, implementing the counter, edge-detection, and data-acquisition functions. A RIGOL DS70304 oscilloscope ($3~\mathrm{GHz}$ bandwidth, $20~\mathrm{GS/s}$ sampling rate) is used to provide reference timing measurements.

\begin{figure}[htbp]
\centering
\includegraphics[width=.47\textwidth]{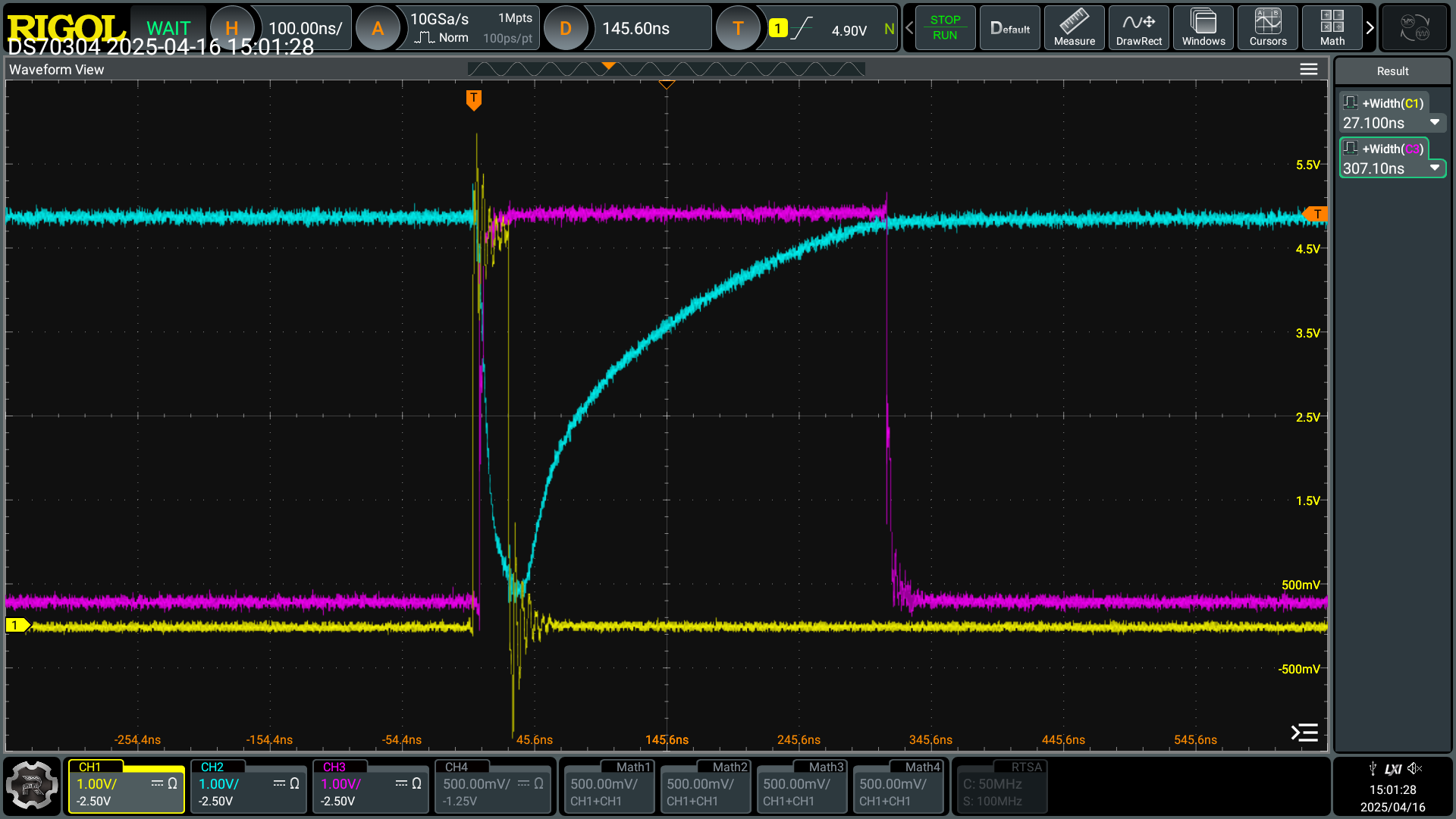}
\qquad
\includegraphics[width=.47\textwidth]{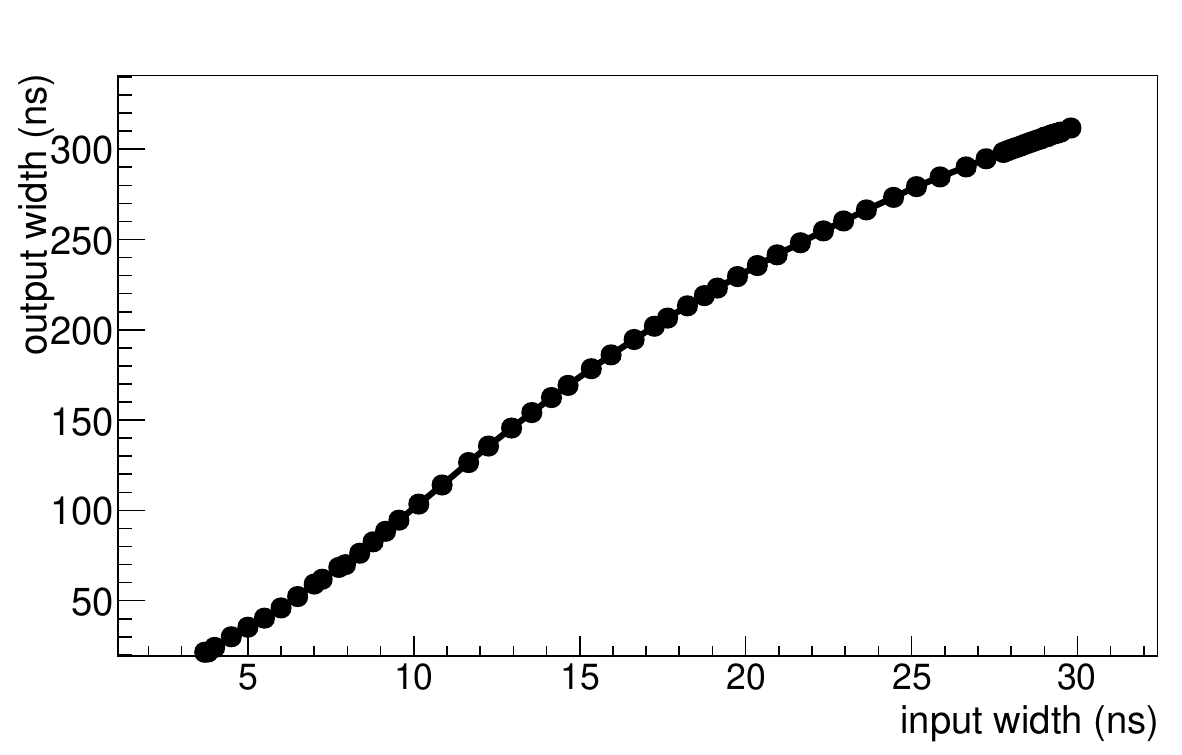}
\caption{Left: Example pulses of vin, vcap and vout of the time-stretching unit measured with an oscilloscope. Right: output pulse width as a function of input pulse width.\label{fig:single_stage_test_pulse}}
\end{figure}

The time-stretching unit was characterized by simultaneously recording its input pulse, capacitor voltage, and stretched output using the oscilloscope. Figure~\ref{fig:single_stage_test_pulse} displays these waveforms for a representative event: the input pulse (yellow), capacitor voltage (blue), and discriminator output (purple, threshold = $4.8~\mathrm{V}$). The plot clearly shows the expected fast discharge and slow charge behavior, though the measured charging profile differs from SPICE simulations (figure~\ref{fig:circuit_singlestage}) - while simulations show constant charging rate, measured data reveals a decreasing rate as voltage approaches maximum. This behavior is due to the behavior of the home-made current source made with LT3092 regulator and current mirrors made with MOSFETs. This non-ideality impacts output linearity, as evidenced in the right plot of figure~\ref{fig:single_stage_test_pulse} showing measured input-output width correlation. The reduced linearity compared to simulation (figure~\ref{fig:simulation_singlestage}) necessitates using interpolation-based look-up tables to determine the stretching factors ($S_0$, $S_1$, $S_2$ in equation~\eqref{eq:4}) for accurate input pulse width conversion.
·~

\begin{figure}[htbp]
\centering
\includegraphics[width=.45\textwidth]{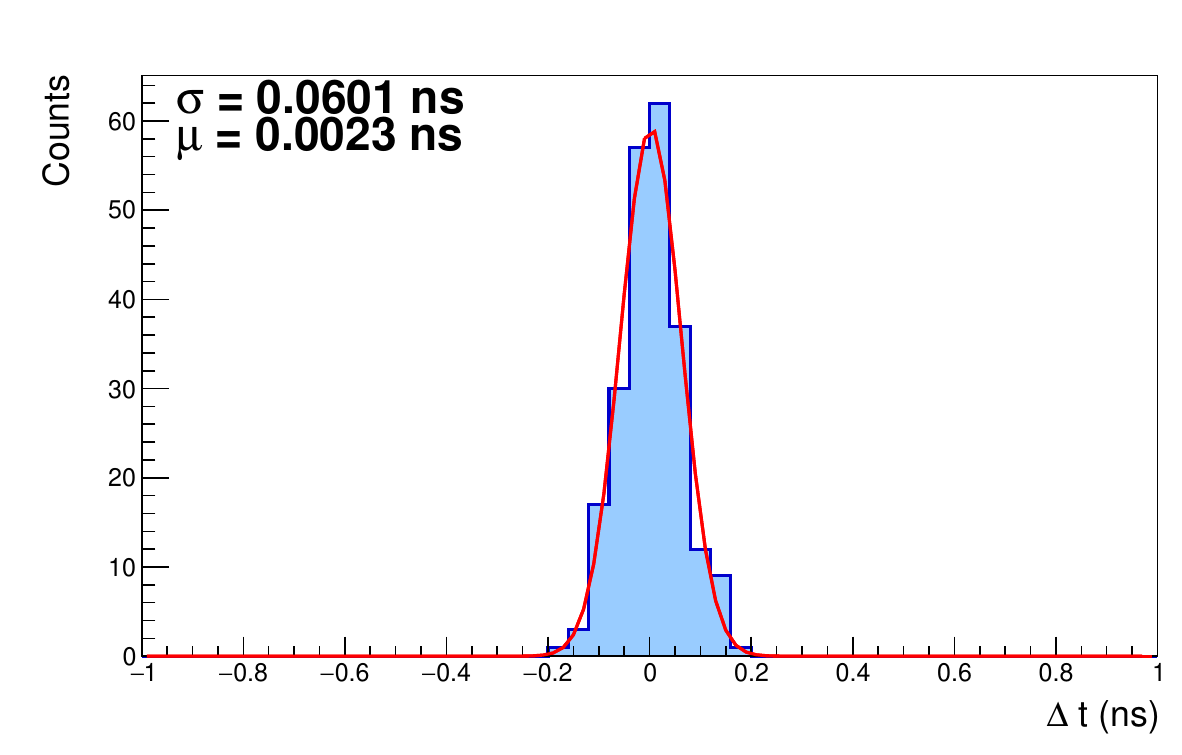}
\qquad
\includegraphics[width=.45\textwidth]{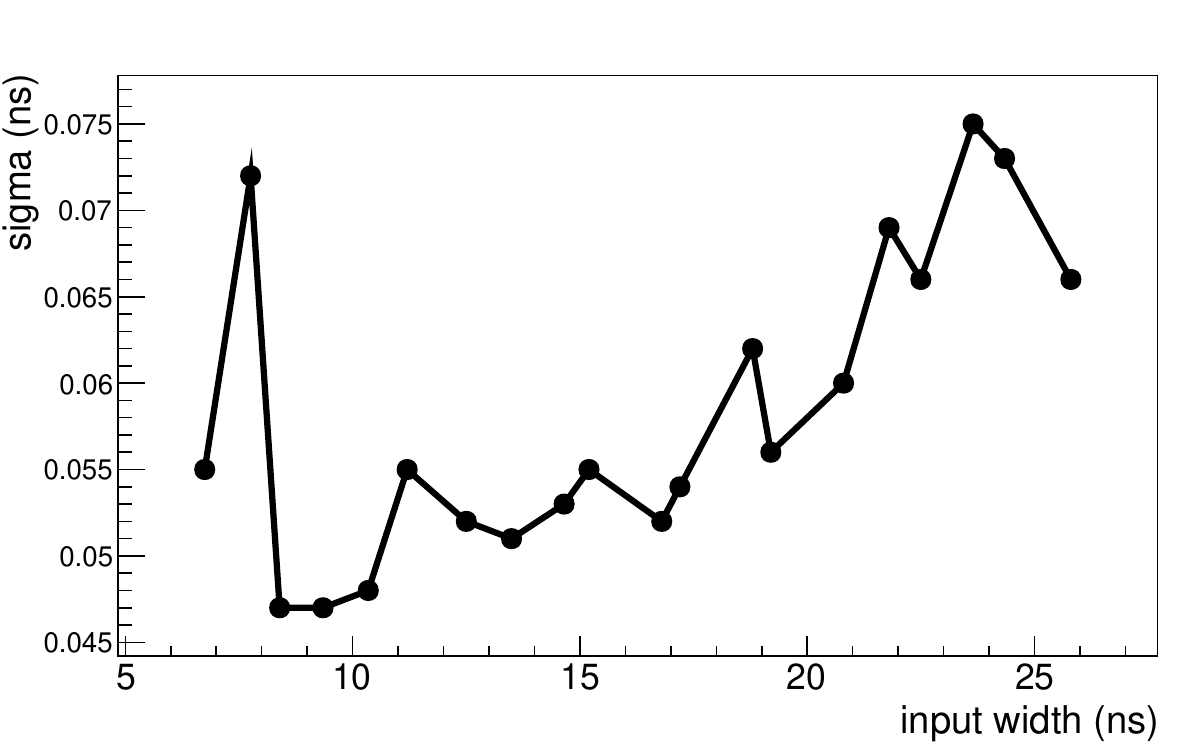}
\caption{Left: distribution of the difference between true pulse width and measured pulse width with time stretching for an input pulse width of 20 ns. Right: time jitter of single-stage time-stretching as a function of the input pulse width.\label{fig:single_stage_time_resolution}}
\end{figure}

The time resolution of the time-stretching TDC has two primary contributions: (1) the LSB quantization error from the counter's finite clock period ($\sigma_t = \mathrm{LSB}/\sqrt{12}$), and (2) additional jitter introduced by the time-stretching process. To characterize the time-stretching jitter, input and output pulse widths were measured using the oscilloscope, with output widths converted to equivalent input widths via the look-up table method. Figure~\ref{fig:single_stage_time_resolution} (left) shows the distribution of differences between measured and true input widths for $\sim\mathrm{20~ns}$ pulses, revealing $\mathrm{60~ps}$ RMS jitter (input-referred) that meets the TDC specifications. The right panel displays jitter versus input width, demonstrating sub-$\mathrm{100~ps}$ performance across the $\mathrm{7}$--$\mathrm{26~ns}$ operational range.

\section{Performance of two-stage time stretching TDC}
\label{sec:perf_two}

\begin{figure}[htbp]
\centering
\includegraphics[width=.45\textwidth]{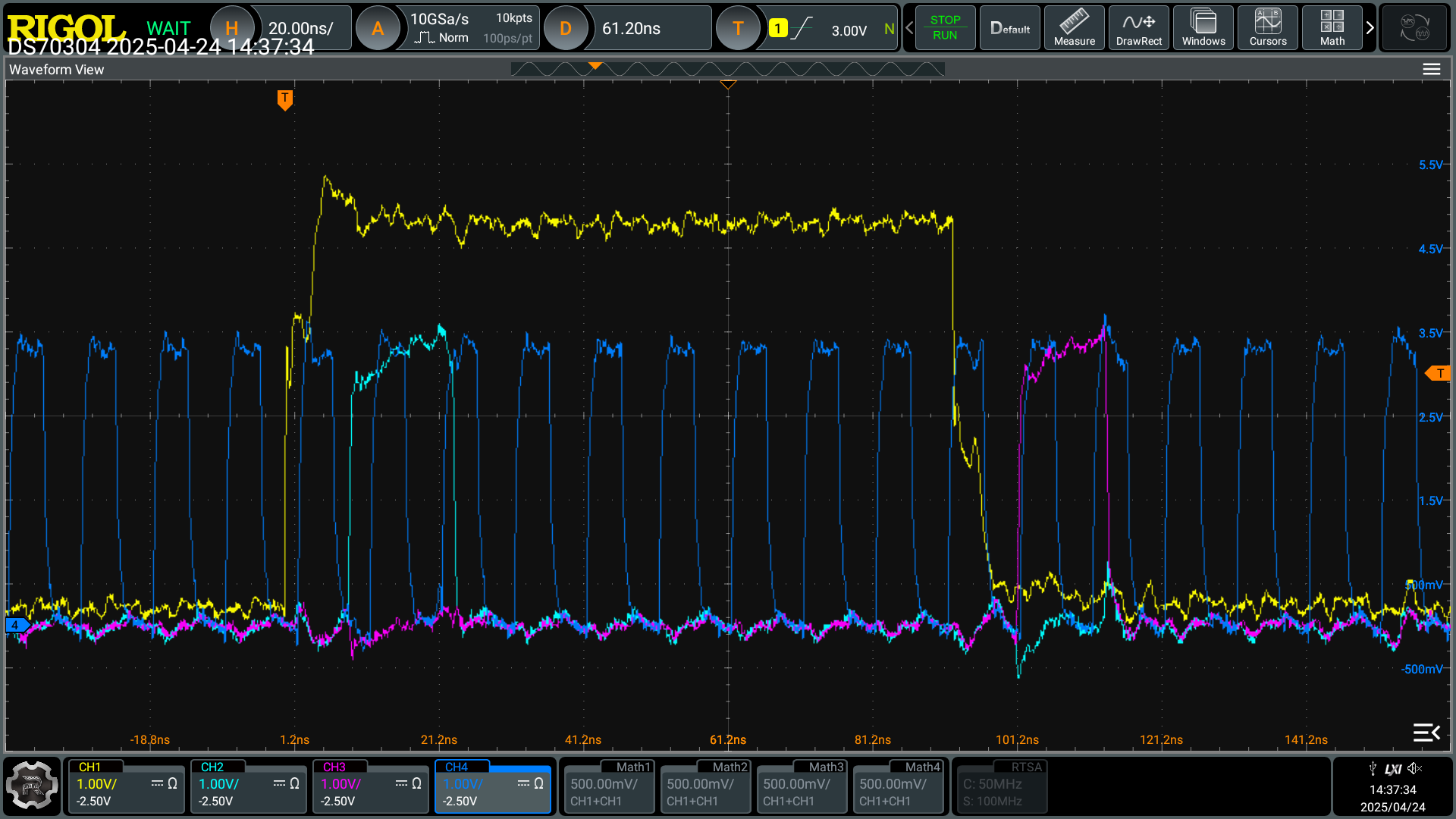}
\qquad
\includegraphics[width=.45\textwidth]{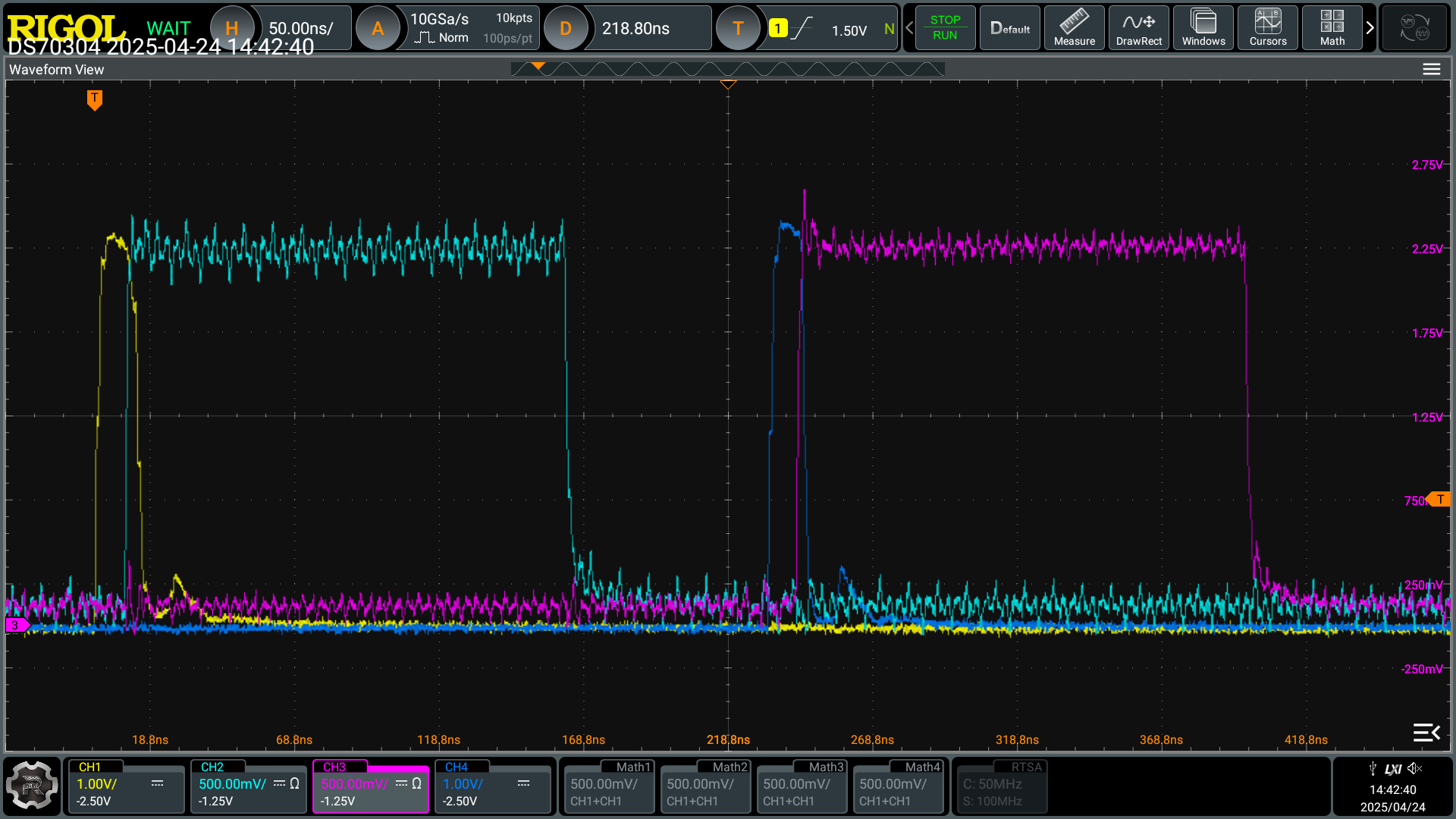}
\caption{Example pulses of the signal chain of the two-stage time-stretching TDC (left: edge detection of the stretched pulse from the output of the first time-stretching stage; right: second stage stretching for the edges.\label{fig:two_stage_test_pulse}}
\end{figure}

Figure~\ref{fig:two_stage_test_pulse} demonstrates the second-stage time-stretching operation, showing both edge detection (left plot) and subsequent stretching (right plot) of the edge1 and edge2 signals defined in section~\ref{sec:intro}. Measurements revealed two systematic effects: (1) a fixed delay between edge-detection outputs and actual signal edges, and (2) a constant offset between detected and true clock-edge distances. These are compensated through a constant correction factor applied to the edge-detection outputs when calculating signal-to-clock-edge distances.

There are two primary uncertainty sources which will affect the edge width measurement, contributing to the TDC's final time resolution:
\begin{enumerate}
    \item \textbf{Second-stage time-stretching units}: Comprising both the time jitter in these units and the LSB error from their counters. These components are similar as the first-stage behavior and are measured separately for edge1 and edge2.
    
    \item \textbf{FPGA edge-detection jitter}: Quantified by comparing the edge-detection output width with the true edge-to-clock distance measured by the oscilloscope.
\end{enumerate}

\begin{figure}[htbp]
\centering

\includegraphics[width=.45\textwidth]{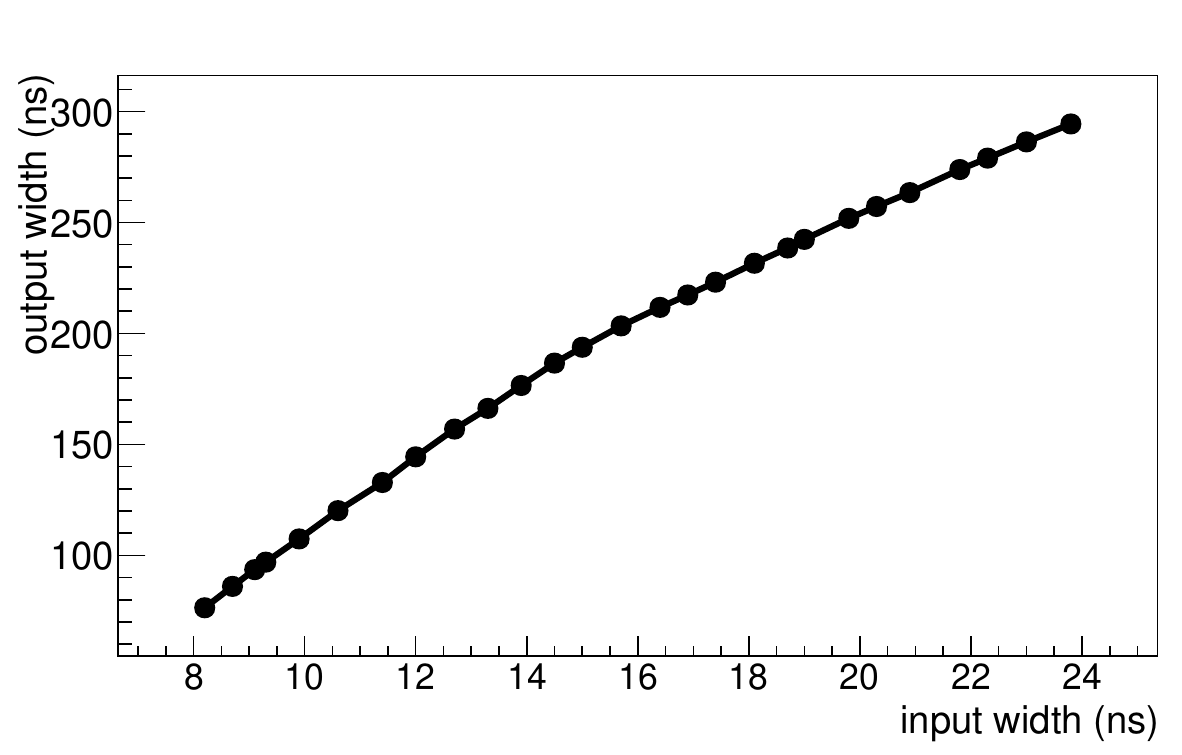}
\qquad
\includegraphics[width=.45\textwidth]{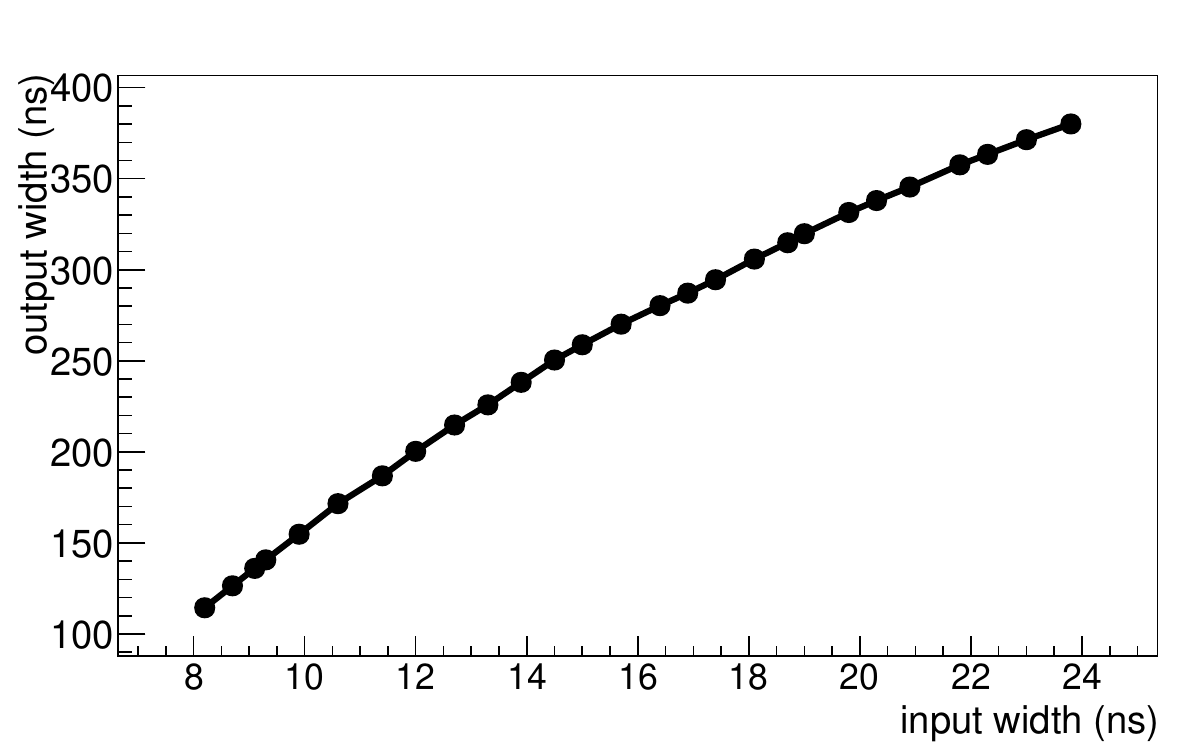}
\caption{Output pulse width as a function of input pulse width for all the two second-stage time-stretching units in the TDC: left is for edge1 and right is for edge2.\label{fig:two_stage_vout_vs_vin}}
\end{figure}

\begin{figure}[htbp]
\centering
\includegraphics[width=.45\textwidth]{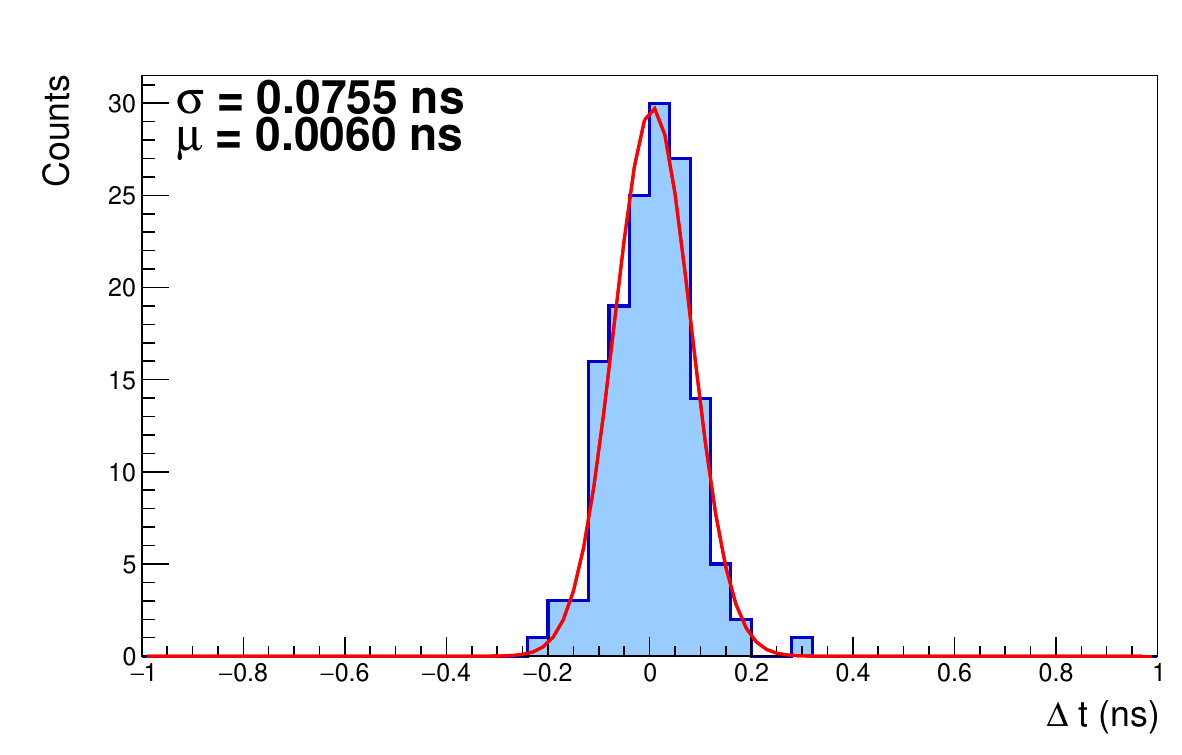}
\qquad
\includegraphics[width=.45\textwidth]{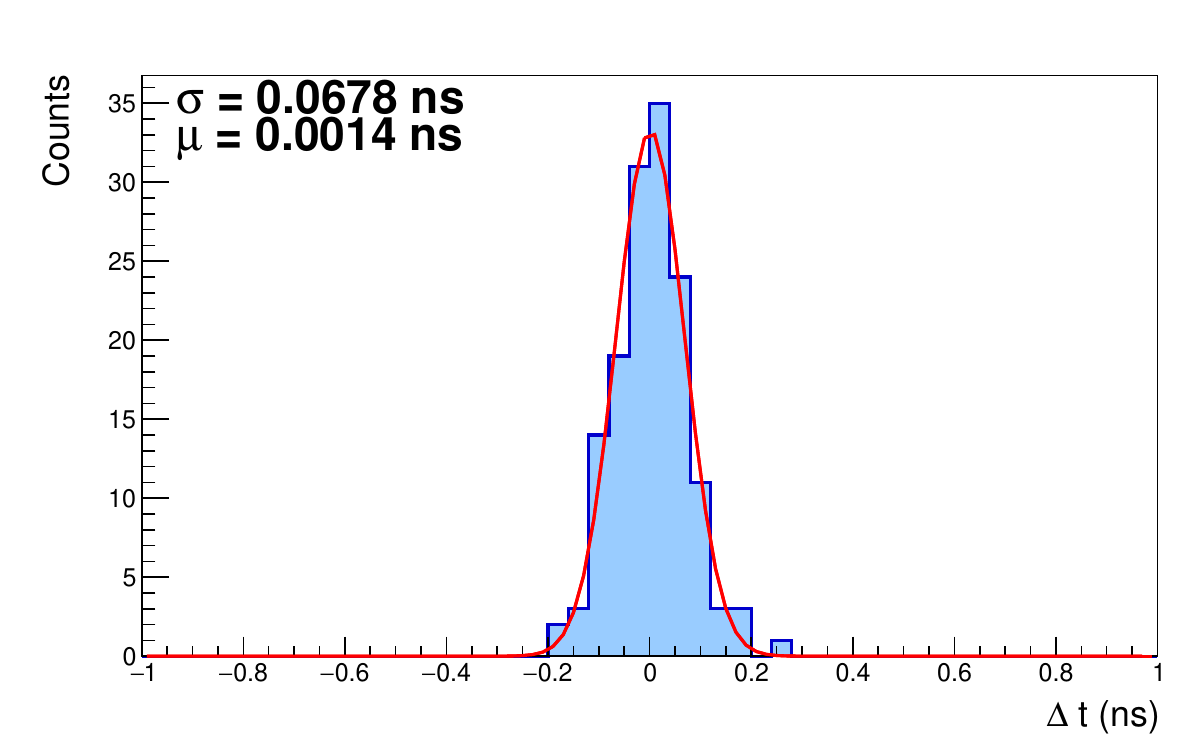}
\caption{Distribution of the difference between true pulse width and measured pulse width with time stretching for the two second-stage time-stretching units in the TDC (true input width is 20 ns): left is for edge1 and right is for edge2.\label{fig:two_stage_time_resolution_gaus}}
\end{figure}

\begin{figure}[htbp]
\centering
\includegraphics[width=.45\textwidth]{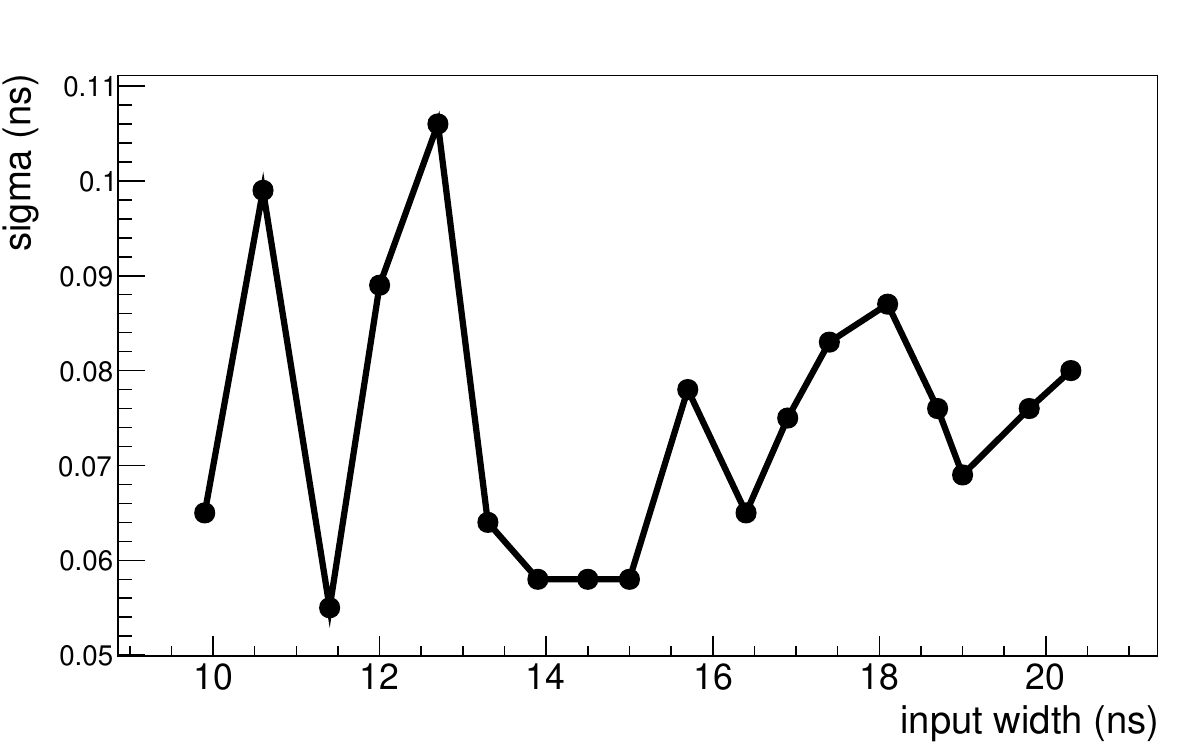}
\qquad
\includegraphics[width=.45\textwidth]{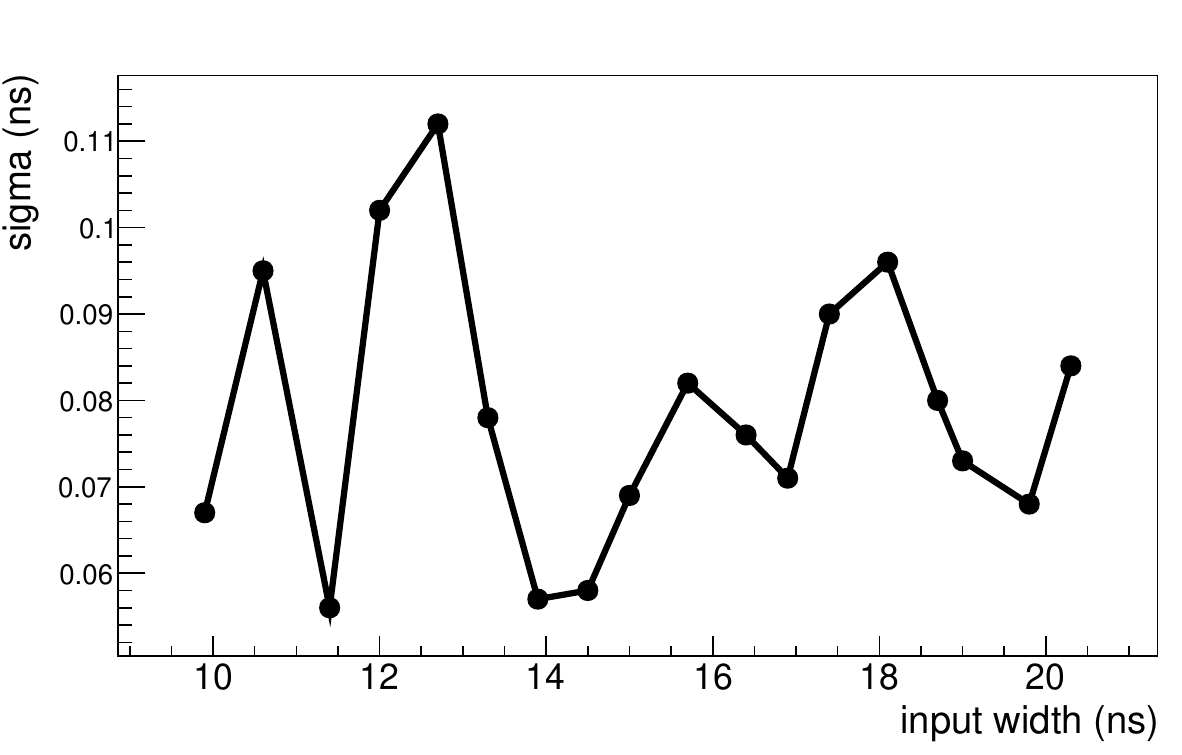}
\caption{Time jitter of the two second-stage time-stretching units as a function of their input pulse width: left is for edge1 and right is for edge2.\label{fig:two_stage_time_resolution_vs_input_width}}
\end{figure}

The second-stage time-stretching units underwent identical characterization as the first-stage unit, with oscilloscope measurements of input/output pulses revealing comparable performance. Figure~\ref{fig:two_stage_vout_vs_vin} displays the output versus input pulse width relationship for both second-stage units, from which look-up tables were generated for pulse width conversion. Figure~\ref{fig:two_stage_time_resolution_gaus} shows the time jitter distribution for a representative $\mathrm{20~ns}$ input pulse, while Figure~\ref{fig:two_stage_time_resolution_vs_input_width} plots jitter versus input width. All measured jitters remain below $120~\mathrm{ps}$, corresponding to $<12~\mathrm{ps}$ input-referred uncertainty when accounting for the first-stage stretching factor.

\begin{figure}[htbp]
\centering
\includegraphics[width=.45\textwidth]{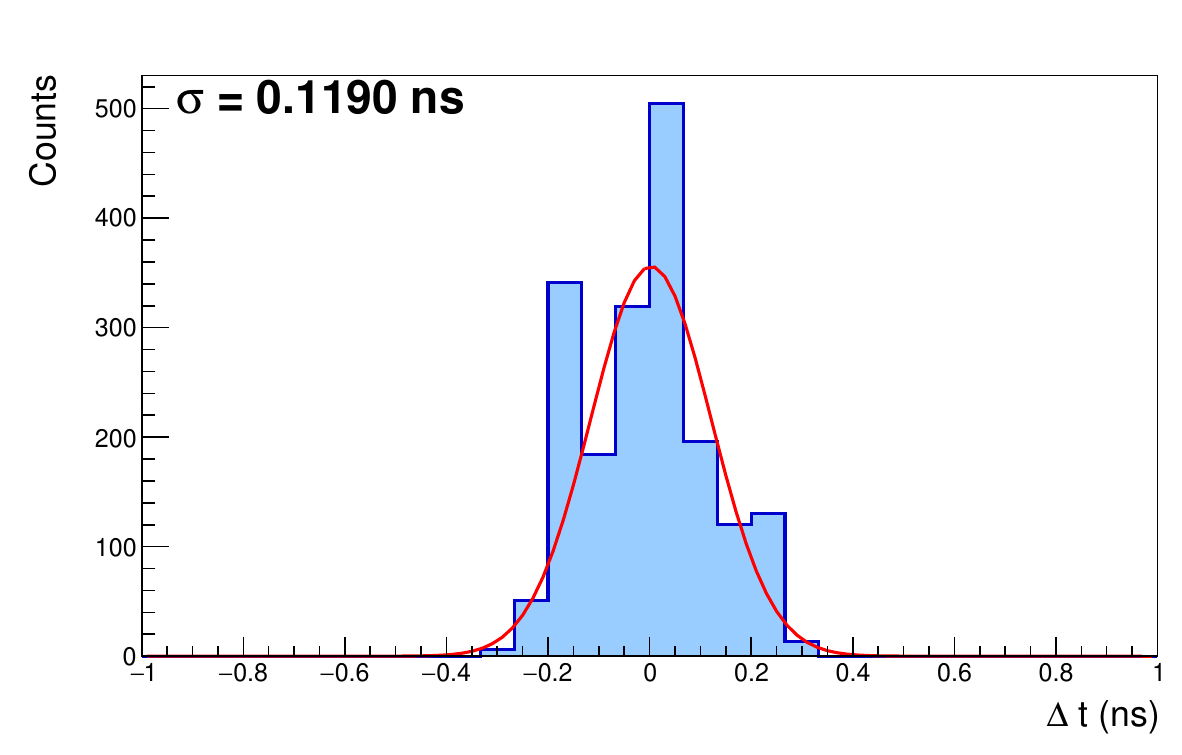}
\qquad
\includegraphics[width=.45\textwidth]{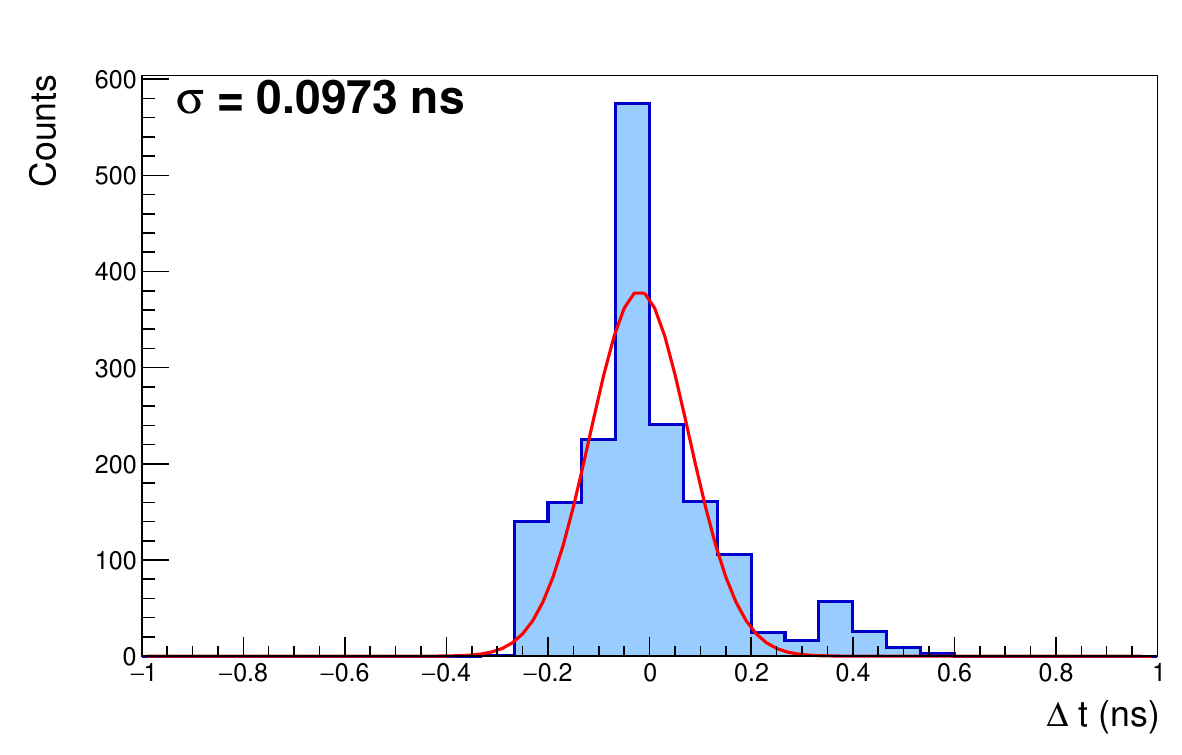}
\caption{Distribution of difference of the width of the edge detection circuit output and the actual distance from the signal edge to clock edge (left: for edge1; right: for edge2).\label{fig:edge_detection_jitter}}
\end{figure}

The FPGA edge-detection jitter was characterized by simultaneously recording three signals with the oscilloscope: (1) the edge-detection input (first-stage stretched pulse), (2) the clock signal, and (3) the edge1/edge2 outputs. By comparing the measured edge1/edge2 widths with the true pulse-to-clock distances, we quantified the detection jitter. Figure~\ref{fig:edge_detection_jitter} shows the distribution of differences between measured and true edge distances, demonstrating FPGA-induced jitter below $\mathrm{120~ps}$. When scaled by the first-stage stretching factor, this contributes $<\mathrm{12~ps}$ to the input pulse width uncertainty.

To sum up, the total time resolution of the two-stage time-stretching TDC combines four independent contributions:
\begin{equation}
    \sigma_t = \sigma_{\mathrm{LSB}} \oplus \sigma_{\mathrm{stretch-jitter-0}} \oplus \sigma_{\mathrm{stretch-jitter-1+2}} \oplus \sigma_{\mathrm{edge-detect-jitter}}
\end{equation}
where:
\begin{itemize}
    \item $\sigma_{\mathrm{LSB}}$: Counter quantization error from the second-stage units. For our $100~\mathrm{MHz}$ clock ($10~\mathrm{ns}$ period), divided by stretching factors $S_0S_1$ and $S_0S_2$, this yields about $100~\mathrm{ps}$ LSBs ($\sigma = 29~\mathrm{ps}$ each). The combined $\sigma_{\mathrm{LSB}} \approx 41~\mathrm{ps}$.
    
    \item $\sigma_{\mathrm{stretch-jitter-0}}$: First-stage time-stretching jitter. Figure~\ref{fig:single_stage_time_resolution} shows $\mathrm{47~ps}$ for $\mathrm{10~ns}$ input pulses.
    
    \item $\sigma_{\mathrm{stretch-jitter-1+2}}$: Second-stage stretching jitters. Figure~\ref{fig:two_stage_time_resolution_vs_input_width} indicates less than $\mathrm{120~ps}$ per unit ($<\mathrm{170~ps}$ combined) for all input width, becoming less than $\mathrm{17~ps}$ input-referred after $S_0$ division.
    
    \item $\sigma_{\mathrm{edge-detect-jitter}}$: FPGA edge-detection jitter. Figure~\ref{fig:edge_detection_jitter} shows $<\mathrm{120~ps}$ per edge ($<\mathrm{170~ps}$ combined), reducing to $<\mathrm{17~ps}$ after $S_0$ division.
\end{itemize}

For $\mathrm{10~ns}$ input pulses, the combined resolution is:
\begin{equation}
    \sigma_t = 41 \oplus 47 \oplus 17 \oplus 17~\mathrm{ps} \approx 67~\mathrm{ps}
\end{equation}

\begin{figure}[htbp]
\centering
\includegraphics[width=.45\textwidth]{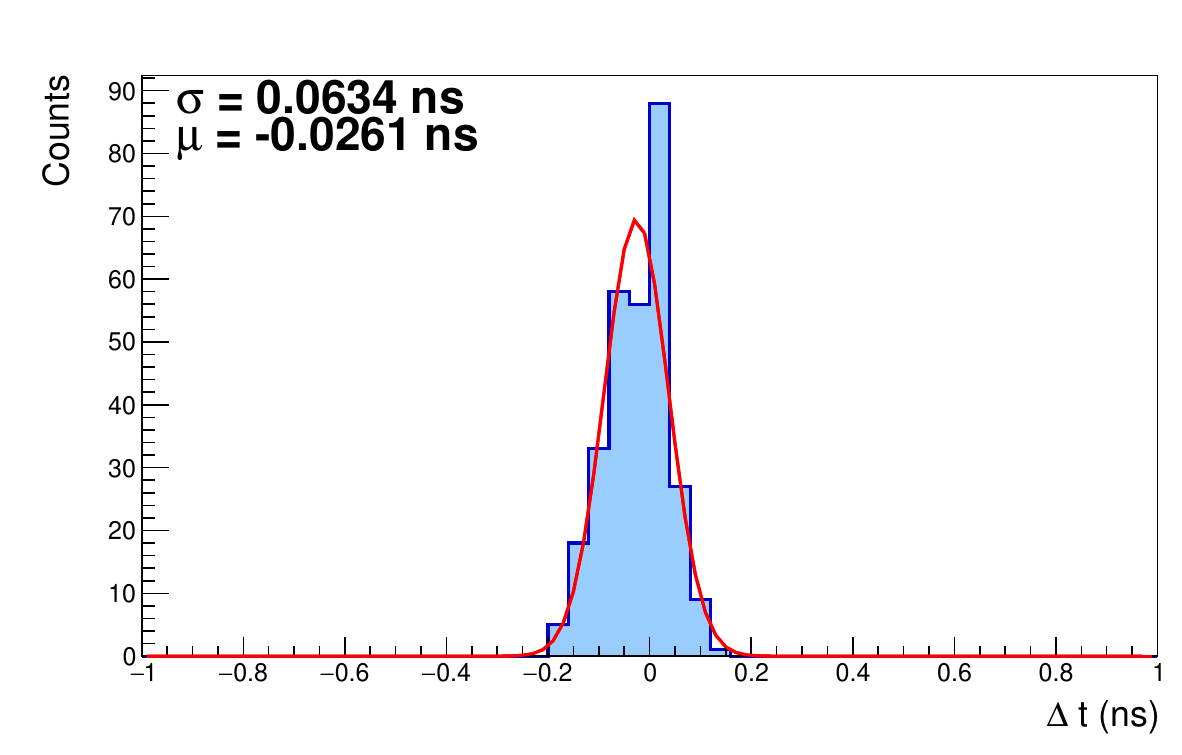}
\qquad
\includegraphics[width=.45\textwidth]{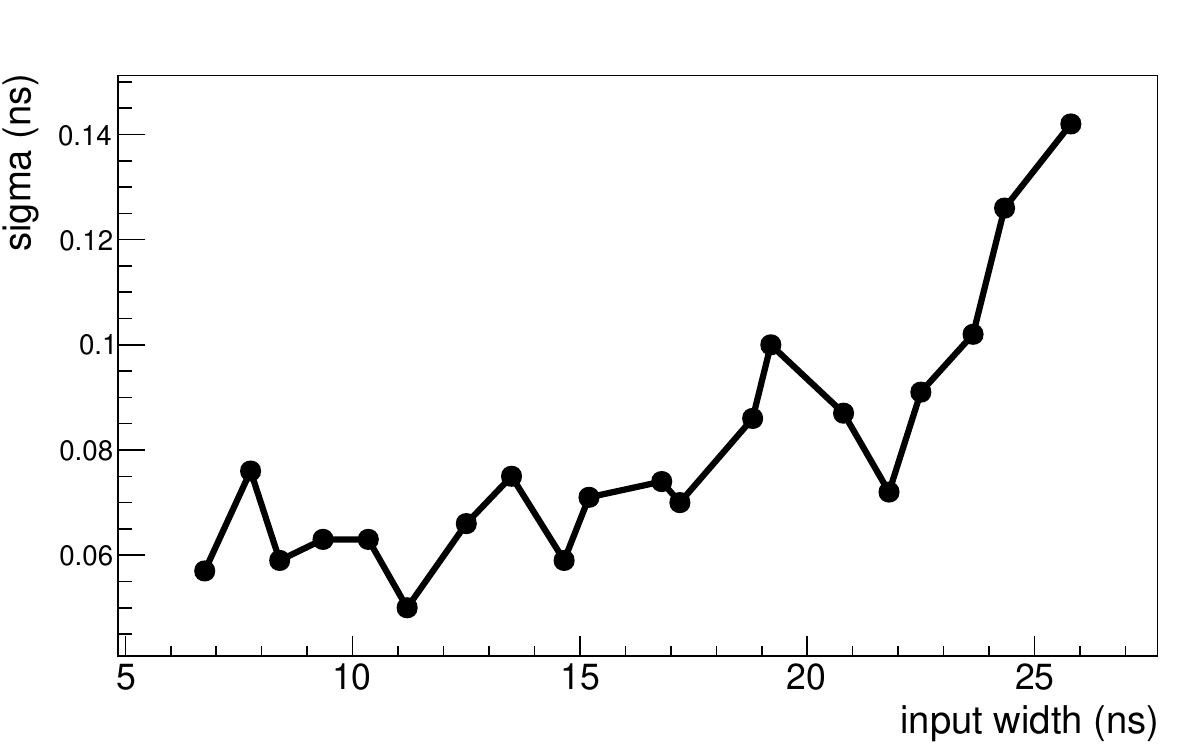}
\caption{Left: Distribution of the difference between measured pulse width and true pulse width for the whole prototype TDC made with two-stage time-stretching (true input width is 10 ns). Right: Time resolution of the prototype TDC as a function of the input pulse width.\label{fig:time_resolution_whole_TDC}}
\end{figure}

The prototype TDC's overall time resolution was measured by comparing direct oscilloscope measurements of input pulse widths with values reconstructed using equation~\eqref{eq:4}. Figure~\ref{fig:time_resolution_whole_TDC} presents these results: the left plot shows the $\Delta t$ distribution for $\sim\mathrm{10~ns}$ input pulses, while the right plot displays resolution versus input width. The measured resolution of $\mathrm{63~ps}$ at $\mathrm{10~ns}$ input width agrees well with our earlier estimate. Across the operational range of $\mathrm{10}$--$\mathrm{20~ns}$ (corresponding to TOA measurements with a $\mathrm{100~MHz}$ clock after adding one clock cycle), the resolution varies between $\mathrm{60}$--$\mathrm{100~ps}$.

\section{Circuit optimization and calibration}
\label{sec:calib}

The discrete-component prototype demonstrates that two-stage time-stretching TDCs can achieve $\mathrm{60}$--$\mathrm{100~ps}$ resolution using only a $\mathrm{100~MHz}$ clock counter. Current performance is primarily limited by first-stage time-stretching jitter (figure~\ref{fig:single_stage_time_resolution}), which future CMOS implementations should significantly reduce.

This prototype serves as an effective platform for optimizing critical design parameters - including clock speeds and stretching factors - prior to integrated circuit fabrication. In the meantime, the demonstrated performance on the prototype made with discrete components already meets requirements for various particle physics applications while providing valuable insights for ASIC development.

The ASIC implementation of this TDC requires careful optimization of two key parameters. First, the clock speed selection involves balancing power consumption (where lower frequencies are preferred) against time resolution requirements from LSB contributions. Second, the stretching factor determination presents a trade-off between improved time resolution (with larger factors) and increased deadtime. Our prototype configuration demonstrates these trade-offs: using a $\mathrm{100~MHz}$ clock and total stretching factor of 100 achieves $\mathrm{41~ps}$ LSB-induced resolution (small compared to the time-stretching unit jitter) while maintaining $\mathrm{300~ns}$ deadtime for $\mathrm{10~ns}$ inputs - suitable for HL-LHC innermost pixel layer rates. The scaling relationships are clearly demonstrated: reducing clock speed by factor $N$ while maintaining resolution requires increasing the total stretching factor by $N$, in the meantime the range for TOA measurement will be increased by a factor of N, resulting in TOA deadtime growth proportional to $N^2$. Our prototype enables rapid validation of these parameter adjustments for specific application requirements.

\begin{figure}[htbp]
\centering
\includegraphics[width=.6\textwidth]{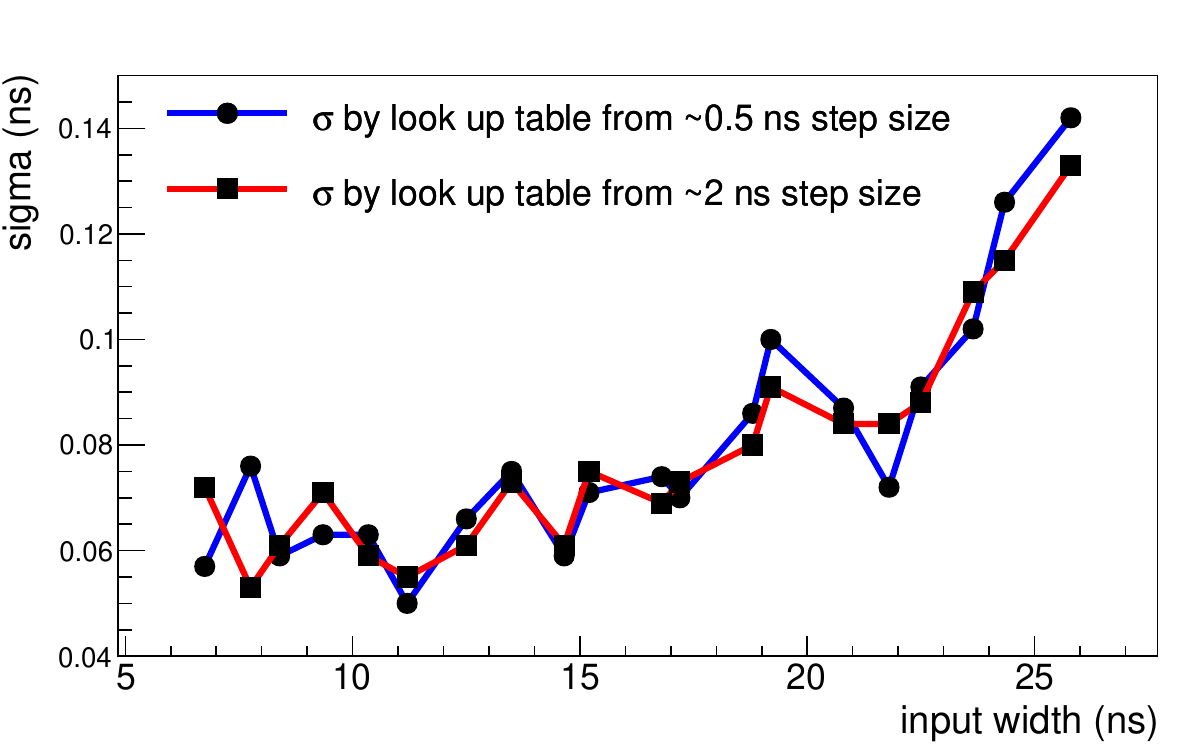}
\caption{Comparison of the time resolution of the prototype TDC when using different numbers of calibration points for the interpolation of the stretching factor look-up table of the time-stretching units.\label{fig:time_resolution_whole_TDC_compare}}
\end{figure}

Calibration is another key aspect of this TDC design, particularly for the time-stretching units. In the current implementation, we determine each unit's stretching factor by injecting input signals and measuring both input and output pulse widths using a high-speed oscilloscope. This process generates calibration plots like figures~\ref{fig:single_stage_test_pulse} and~\ref{fig:two_stage_vout_vs_vin}. The input-output width relationship shows noticeable non-linearity, requiring multiple data points and interpolation to create accurate look-up tables for converting output widths to input widths. While figures~\ref{fig:single_stage_test_pulse} and~\ref{fig:two_stage_vout_vs_vin} contain many data points (about $0.5~\mathrm{ns}$ step size) for this interpolation, we find similar performance can be achieved with far fewer calibration points. Specifically, figure~\ref{fig:time_resolution_whole_TDC_compare} demonstrates that using calibration points of $2~\mathrm{ns}$ step size yields time resolution comparable to our results obtained from calibration points of $0.5~\mathrm{ns}$ step size.

The calibration method described above works well for single-channel implementations like our prototype, but becomes impractical for multi-channel ASIC designs required in pixel detector readout systems. For such applications, a more practical calibration can be achieved by injecting precisely known pulse widths into each time-stretching unit and measuring the outputs using a dedicated calibration TDC integrated on the ASIC. Both the pulse injection circuitry and calibration TDC can be shared across multiple channels to conserve resources. The test pulses can originate either from an on-chip pulse generator similar to our prototype PCB design, or from single cycles of a configurable clock source using clock division or PLL techniques. The dedicated calibration TDC itself can be implemented as a simple high-speed counter, providing an area-efficient solution suitable for large-scale ASIC implementations with many TDC channels.

To study the requirement for the precision of the dedicated calibration TDC, a toy Monte-Carlo (MC) study was performed on a mock-up two-stage time-stretching TDC. 

\begin{figure}[htbp]
\centering
\includegraphics[width=.45\textwidth]{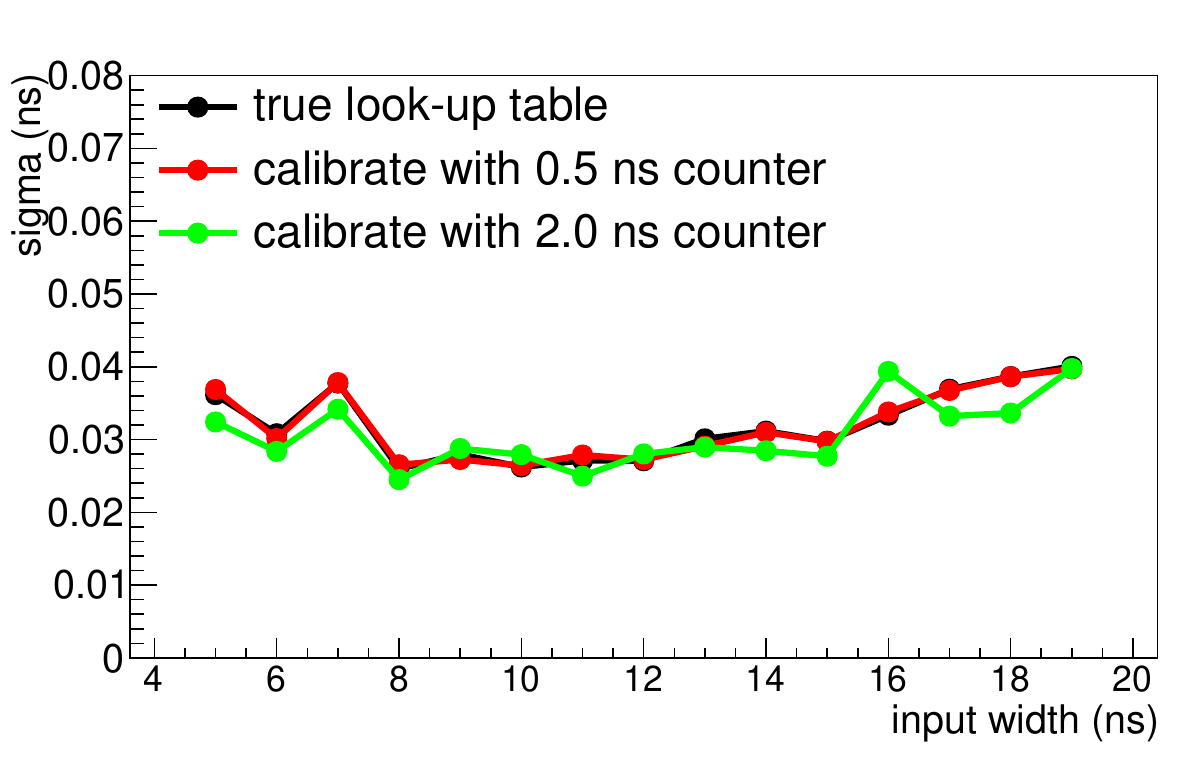}
\qquad
\includegraphics[width=.45\textwidth]{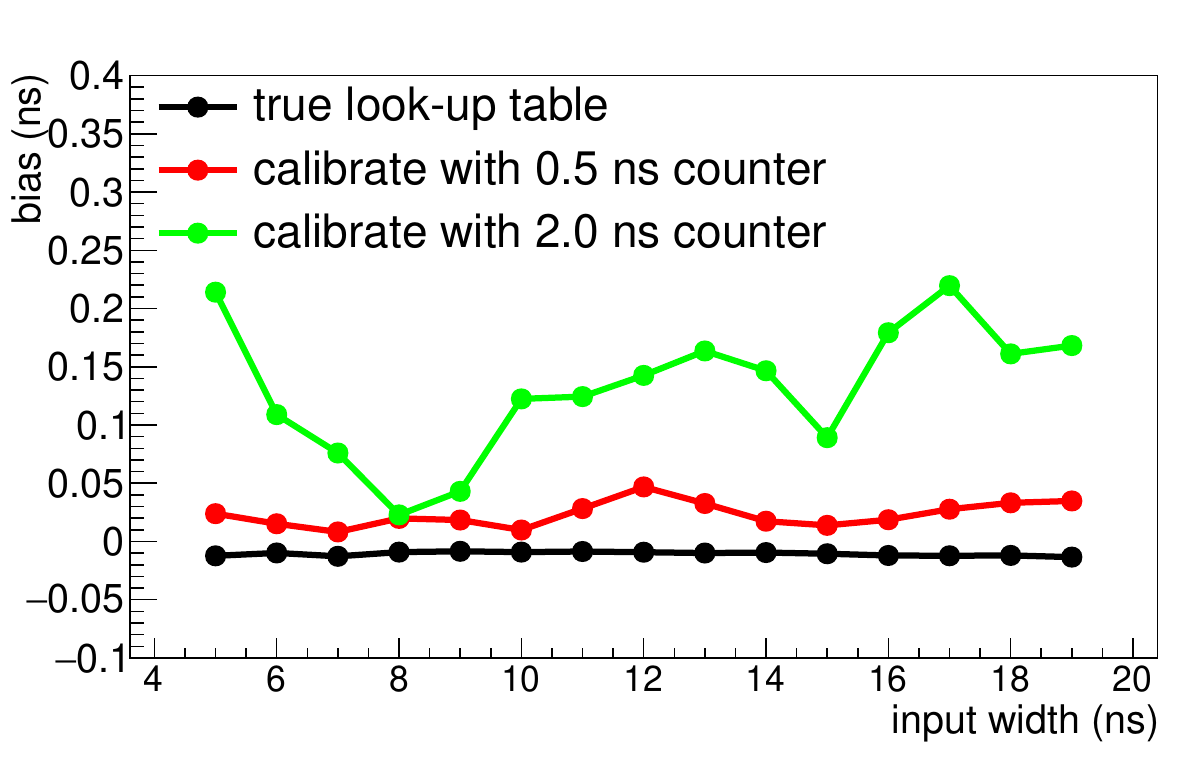}
\caption{Comparison of TDC performance with calibration of the stretching factor measured with counters of different speed: the left plot is the time resolution of the TDC and the right plot is the time measurement bias of the TDC.\label{fig:calibration_with_different_clock}}
\end{figure}

First, the contribution of counter LSB ($\mathrm{10~ns}$) to the TDC performance was simulated by injecting signals of known width and measuring $N_0$, $N_1$, and $N_2$ in the toy MC. These values were converted to measured input width based on stretching factors from look-up tables derived from figures~\ref{fig:single_stage_test_pulse} and~\ref{fig:two_stage_vout_vs_vin} (using about $\mathrm{1~ns}$ step size for calibration points). The black points in figure~\ref{fig:calibration_with_different_clock} show the resulting resolution and bias of the time measurement versus input width. The observed resolution of $\sim\mathrm{40~ps}$ matches our LSB contribution estimate from section~\ref{sec:perf_two}.

Next, instead of using ideal look-up tables, we calibrated the stretching factors by injecting known-width signals into each time-stretching unit and measuring the outputs with realistic dedicated calibration TDCs (LSBs of $\mathrm{0.5~ns}$ and $\mathrm{2~ns}$ were tested). This generated calibration curves similar to figures~\ref{fig:single_stage_test_pulse} and~\ref{fig:two_stage_vout_vs_vin}, from which new stretching factor look-up tables were created. The red and green points in figure~\ref{fig:calibration_with_different_clock} show the performance using these calibrations with $\mathrm{0.5~ns}$ and $\mathrm{2~ns}$ LSB calibration TDCs, respectively.

The results show that calibration TDC precision has minor impact on resolution but introduces additional measurement bias. A $\mathrm{0.5~ns}$ LSB calibration TDC gives negligible effect on both resolution and bias, while a $\mathrm{2~ns}$ LSB version remains viable if the bias can be determined from physics events, such as pairs of high-energy particles from the same vertex in collider experiments.

\section{Summary}
\label{sec:summary}

In summary, the design and test results of a two-stage time-stretching prototype TDC made with low-cost discrete components are presented. With about a factor of 10 time stretching at each stage, the effective LSB is reduced by about a factor of 100. A less than 100 ps time resolution is achieved with a 100 MHz clock counter. The time resolution of such a TDC is found to be dominated by the time jitter of the first-stage stretching unit. The conversion time for a 10 ns input width is less than 300 ns, a factor of five less than a single stage stretching with a stretching factor of 100. The prototype TDC presented in this paper is a stepping stone and optimization platform towards the design of TDC in modern CMOS technologies for future 4D pixel detectors at hadron colliders, which requires $\mathrm{\mu W}$ level power consumption, $\mathrm{<50~ps}$ level time resolution, and with more than $\mathrm{100~kHz}$ hit rate. The prototype is also a ready-to-use low-cost system for many existing applications where a $\mathrm{100~ps}$ level resolution is enough, such as precision timing circuits for neutrino detectors.


\acknowledgments

This work was supported by Tsinghua University Initiative Scientific Research Program.


\bibliographystyle{JHEP}
\bibliography{main.bib}

\end{document}